\documentclass[11pt,a4paper]{article}

\pdfoutput=1
\usepackage{jheppub}

\usepackage{amssymb}
\usepackage{amsmath}
\usepackage{amsfonts}
\usepackage{graphicx}
\usepackage{color}
\usepackage{xspace}
\usepackage{ulem}
\usepackage{mathtools}
\usepackage{hhline}
\usepackage{float}

\begin{document}

\title{Determining the spin of light primordial black holes with Hawking radiation}

\author[a]{Marco Calz\`a}    \emailAdd{mc@student.uc.pt}
\author[a]{Jo\~{a}o G.~Rosa} \emailAdd{jgrosa@uc.pt}

\affiliation[a]{Univ Coimbra, Faculdade de Ci\^encias e Tecnologia da Universidade de Coimbra and CFisUC, Rua Larga, 3004-516 Coimbra, Portugal}

%\newline\newline
\abstract{
   We propose a method to determine the mass and spin of primordial black holes (PBHs) in the mass range $5\times 10^7-10^{12}$ kg (Hawking temperatures  $\sim10$ MeV $-200$ GeV), based on measuring the energy of specific features in the photon Hawking emission spectrum, including both primary and secondary components. This is motivated by scenarios where PBHs in this mass range spin up as they evaporate, namely the string axiverse, where dimensionless spin parameters $\tilde{a}\sim 0.1-0.5$ can be achieved through the Hawking emission of hundreds or even thousands of light axion-like particles. Measuring the present PBH mass-spin distribution may thus be an important probe of physics beyond the Standard Model. Since the proposed method relies on the energy of the photons emitted by a given PBH, rather than on the associated flux, it is independent of the PBH-Earth distance and, as a byproduct, can also be used to infer the latter.}

   % We propose a distance independent method to characterize the mass and the spin of asteroids mass black holes through the observation of their photon spectrum. We take into account non-nearly-extremal BHs with $a_* \in 0\div0.5$ and masses between $10^{12} \div 5 \times 10^7$ kg. We provide simple relations conneticng the black hole mass and spin with energies and intensities of specific structure in the total photon spectrum. According to the mass range of the black hole we took into account appropriate methods for obtaining the black hole spectrum. As a by product we derive a method to measure the distance of a black hole based on the measure of its photon intensity/flux.

\maketitle

\section{Introduction}

There has been a growing interest in the literature on the study of primordial black holes (PBHs) formed in the early Universe through the gravitational collapse of putative overdense regions \cite{Hawking:1971ei, Carr:1974nx, Carr:1975qj}. Several mechanisms for PBH production have been proposed in the literature, within a broad range of possible masses (see e.g. \cite{Carr:2020xqk}). Depending on their mass, PBHs may account for a significant fraction of the present dark matter abundance and potentially explain the low spins of the merging BH binaries detected by LIGO/Virgo \cite{Clesse:2017bsw, Sasaki:2016jop} or even the ultrashort timescale microlensing events observed by OGLE in recent years \cite{Niikura:2019kqi}.

Light PBHs with mass $\lesssim 10^{12}$ kg are particularly interesting since they should be evaporating today, following Hawking's original proposal \cite{PartCrea}. Their photon Hawking emission contributes to the extra-galactic gamma-ray background, from which strong constraints on their contribution to dark matter have been placed in recent years \cite{Carr:2009jm, Carr:2016hva, Carr:2020gox, Arbey:2019vqx, Ferraz:2020zgi, Auffinger:2022khh}. Nevertheless, even if such light PBHs only account for a tiny fraction of the dark matter density, $f\lesssim 10^{-7}$ \cite{foot1}, there could still be numerous PBHs in our astrophysical neighbourhood, where $\rho_{DM}\simeq 2\times10^{12}\ \mathrm{kg}\,\mathrm{AU}^{-3}$, within distances as low as $\lesssim 100$ AU. The possibility of observing such PBHs and testing Hawking's prediction of a nearly-thermal emission spectrum has motivated searches for evaporating PBHs with gamma-ray instruments such as H.E.S.S. \cite{Glicenstein:2013vha, Tavernier:2019exh}, HAWC \cite{HAWC:2019wla}, Milagro \cite{Abdo:2014apa}, VERITAS \cite{Archambault:2017asc} and Fermi-LAT \cite{Fermi-LAT:2018pfs} (see also \cite{Halzen:1990ip, Halzen:1991uw, Ukwatta:2009xk, MacGibbon:2015mya}), although so far none have been found.

A standard lore in the literature has been that PBHs spin down as they evaporate, which is true if only the known Standard Model (SM) particles are emitted through Hawking evaporation, since all emitted particles with non-zero spin necessarily carry away a part of the PBH's angular momentum \cite{Page:1976df, Page:1976ki}. Scalar particles, however, can be emitted in the spherical monopole mode ($l=0$), such that their emission reduces the mass $M$ but not the spin $J$ of a PBH, therefore increasing the dimensionless spin parameter $\tilde{a}= J M_P^2/M^2$, where $M_P\simeq 2.17\times10^{-8}$ kg denotes the Planck mass. This was originally found by Taylor, Chambers and Hiscock (TCH) \cite{Chambers:1997ai, Taylor:1998dk}, who showed that a BH evaporating through pure scalar emission would asymptote to a value $\tilde{a}\simeq 0.555$ towards the end of its lifetime. In the SM, the number of scalar degrees of freedom that can be emitted by a PBH is completely overwhelmed by the number of particles with non-vanishing spin, taking into account that pions are the only composite mesons that can be directly emitted by a PBH, since for Hawking temperatures above the QCD scale free quarks are emitted instead \cite{MacGibbon:1990zk, MacGibbon:1991tj} (see also \cite{MacGibbon:1991vc, MacGibbon:2007yq, MacGibbon:2010nt, Ukwatta:2015iba}), and that Higgs bosons can only be significantly emitted for Hawking temperatures above the electroweak scale. As a result, within the SM a PBH always spins down completely even before losing a significant fraction of its mass.

This, however, is no longer true if one considers beyond the SM (BSM) scenarios with many light scalars ($\lesssim$ few MeV), as we have recently shown with March-Russell in \cite{Calza:2021czr}. This is the case of the string axiverse proposal \cite{Arvanitaki:2009fg}, which argues that realistic string theory compactifications result in a large number of light axion-like particles (ALPs), typically $\mathcal{O}(100-1000)$ or possibly an even larger number. These ALPs are associated with the Kaluza-Klein zero-modes of higher-dimensional gauge fields, like the Neveu-Schwarz or Ramond-Ramond $p$-forms, and the large number of non-trivial compact cycles that can support these in a 6-dimensional compact (Calabi-Yau) manifold. By realistic we mean a scenario that admits at least one light axion, the QCD axion, associated with the Peccei-Quinn solution to the strong CP-problem \cite{Peccei:1977hh, Wilczek:1977pj}. A light axion can only exist if supersymmetry breaking or moduli stabilization mechanisms do not break axion shift symmetries, which implies that most ALPs only acquire masses through non-perturbative effects and therefore remain light. As we have shown in \cite{Calza:2021czr}, adding $\gtrsim 100$ light axions to the SM can significantly slow down a PBH's spin down as it evaporates or even spin it up, such that PBHs with mass $\lesssim 10^{12}$ kg may presently rotate with $\tilde{a}\gtrsim 0.1$, and even close to the upper bound found by TCH for $\gtrsim 10^3$ light ALP species.

In the context of the string axiverse, and potentially other SM extensions with large numbers of light scalars, evaporating PBHs should presently exhibit a non-trivial spin, and measuring the present mass-spin distribution of PBHs at different stages of their lifetime may therefore yield a unique probe of new physics. In our previous work we showed that the spectrum of primary photons (those directly emitted by the PBH) is sensitive to both the mass and spin of a PBH, such that detecting such photons could allow one to determine both quantities. This, however, relies on the measured photon flux, and hence depends also on precise measurements of the PBH-Earth distance. While this may be relatively easy to do with parallax techniques if a PBH is not too far away (which is nevertheless required for detection), it would be better to devise a mass-spin determination method independent of distance measurements. In addition, the observed flux will correspond not only to the primarily emitted photons but also to secondary photons radiated by the other particles that also result from the PBH's evaporation. These result from final state radiation (FSR) of charged particles, particle decays (muons, pions, etc) as well as parton fragmentation processes, and can be more numerous, albeit less energetic, than the primary photons. 

In this article, we therefore study the full Hawking photon spectra of PBHs with temperatures roughly between 10 MeV and 200 GeV, corresponding to PBH masses from $10^{12}$ kg down to $5\times 10^{7}$ kg, respectively, which are significantly evaporating today but are sufficiently cold for us to ignore any contributions from potential new particles with masses above the electroweak scale (see e.g. \cite{Baker:2021btk, Baker:2022rkn} for the effects of new heavy particles on PBH evaporation). We will also ignore photons from ALP decays, since string axions typically have large decay constants (above the GUT scale) and therefore very long lifetimes/small decay rates. We employ the publicly available BlackHawk code \cite{Arbey:2019mbc, Arbey:2020yzj, Arbey:2021yke, Arbey:2021mbl} to compute the Hawking spectra of PBHs with different mass and spin, which uses both PYTHIA \cite{Sjostrand:2007gs, Bierlich:2022pfr} and Hazma \cite{Coogan:2019qpu} as particle physics codes to determine the secondary photon emission, taking into account the regime of validity of the latter codes. For the heavier (and hence colder) PBHs that cannot effectively emit particles heavier than the electron, we employ semi-analytical tools to determine the secondary photon spectrum resulting mainly from electron FSR following \cite{Coogan:2020tuf}. 

We identify particular features in the Hawking spectrum and from their energies (and energy ratios) devise a method to determine both the mass and spin of a PBH in the above-mentioned mass range. We will show, in particular, that, with this methodology, one can distinguish an effectively Schwarzschild PBH from one that is spinning with $\tilde{a}\simeq 0.1-0.5$ as the result of light scalar emission. By using the energy of these features rather than the corresponding photon flux, the method is thus independent of the distance between the observer and the PBH. One can then use the inferred PBH mass and spin to predict the expected photon flux as a function of the Earth-PBH distance, therefore allowing one to determine the latter from the measured flux, alongside potential parallax measurements.

This work is organized as follows. Section II contains a brief review of the numerical and semi-analytical methods used in calculating the primary and secondary PBH Hawking spectra. In this section we also discuss the approximations used and the range of validity of each method. In Section III we describe the general characteristics of the photon spectrum, discussing the differences observed for distinct PBH masses and spins. In section IV we outline our proposed method to determine the PBH mass and spin and discuss the energy resolution requirements for sufficiently precise measurements of these quantities that may allow one to detect the effects of new physics, particularly those of the string axiverse. We summarize our main conclusions and discuss future prospects in Section V.

%%%%%%%%%%%%%%%%%%%%%%%%%%%%%%%%%%%%%%%%%%%%%%%%%%%%%%%%%%%%%%%%%%%%%%%%%%%
%%%%%%%%%%%%%%%%%%%%%%%%%%%%%%%%%%%%%%%%%%%%%%%%%%%%%%%%%%%%%%%%%%%%%%%%%%%
%%%%%%%%%%%%%%%%%%%%%%%%%%%%%%%%%%%%%%%%%%%%%%%%%%%%%%%%%%%%%%%%%%%%%%%%%%%
%%%%%%%%%%%%%%%%%%%%%%%%%%%%%%%%%%%%%%%%%%%%%%%%%%%%%%%%%%%%%%%%%%%%%%%%%%%

\section{Computing the Hawking spectrum}

In this section we outline the methodologies adopted for computing the primary and secondary components of the photon spectrum.
We have considered different methods according to the energies of the emitted particles and therefore to the mass of the PBH. We have, in particular, used the BlackHawk code to determine the secondary photon spectrum, coupled with the Hazma/PYTHIA codes for low/high energy primary particles. For low-temperature PBHs ($\lesssim 20$ MeV), we employ the semi-analytical method described in \cite{Coogan:2020tuf} to compute the secondary spectrum due to the electron's FSR, which is the dominant contribution. These methods rely on different approximations and have different ranges of validity, which we discuss in each case.

\subsection{Primary spectrum}

A BH is characterized by an outgoing flux of particles. This is due to the different quantum vacuum definitions of a near horizon free-falling observer and a distant observer at rest in the BH frame. As a result a net nearly-thermal flux at infinity emerges. For a Kerr BH with dimensionless spin parameter $\tilde{a}$ and angular velocity at the horizon $\Omega$, a particle species $i$ with spin $s$ is emitted at a differential rate \cite{PartCrea, Page:1976df, Page:1976ki}:
\begin{equation}\label{prim}
{d^2N_{P, i}\over dt dE_i}={1\over 2\pi}\sum_{l,m}{\Gamma^s_{l,m}(\omega)\over e^{2\pi k/\kappa}\pm 1}~,
\end{equation}
where $\omega=E_i$ is the mode frequency (in natural units), $\Gamma^s_{l,m}$ are the absorption coefficients or gray-body factors encoding the deviations from a black-body spectrum for each $(l,m)$-mode in a spheroidal wave decomposition, $k=\omega-m\Omega$ and $\kappa=\sqrt{1-\tilde{a}^2}/2r_+$ is the surface gravity of the Kerr BH with event horizon at $r_+$. Note that the plus (minus) sign in the denominator corresponds to fermions (bosons). We calculated $\Gamma^s_{l,m}$ by solving the Teukolsky equation \cite{Teukolsky:1972my, Teukolsky:1973, Press:1973zz, Teukolsky:1974yv} that describes wave scattering in the Kerr geometry up to $l=4$ for photons ($s=1$) using a shooting method (see e.g.~\cite{Rosa:2016bli}) A similar procedure can be applied in computing the spectrum of all other primary particles contributing to the secondary photon spectrum, and we have considered up to $l=4$ ($l=9/2$) modes for bosons (fermions) in our calculations, in the cases where we explicitly compute the primary spectrum of the emitted particles as discussed below. We note that these upper bounds on $l$ are sufficient to accurately compute the primary spectra for the values $\tilde{a}\lesssim 0.5$ that we are interested in, whereas for larger values of the spin parameter higher-$l$ modes must be included in the computation.

\subsection{Secondary spectrum}

An evaporating PBH emits several different charged particles that radiate photons as they travel away from the PBH. Photons also result from the decay of unstable particles, like neutral pions. These photons are naturally less energetic than those emitted directly by the Hawking effect, but nevertheless yield a very significant contribution to the total photon spectrum, in some cases a few orders of magnitude above the intensity of primary emission. The full spectrum can then be obtained by convoluting the primary emission rate in Eq.~(\ref{prim}) with the number of photons radiated by each charged/unstable primary particle. This is, generically, a non-trivial procedure that has to be performed using numerical tools, particularly in the case of quarks and gluons that hadronize as they move away from the PBH, for Hawking temperatures roughly exceeding $\Lambda_{QCD}$. 

We have chosen to use the publicly available BlackHawk code for numerical calculations of the secondary spectrum. In fact, this code gives the full photon spectrum for a given PBH mass and spin, and we have checked that the primary component yields results in agreement with our independent calculation. We must note that the latest version of BlackHawk relies on two different particle physics codes to compute the number of photons radiated by primary particles: Hazma for primary particle energies below a few GeV and PYTHIA for energies $>5$ GeV.

Let us note that the PBH primary emission has a nearly-blackbody shape, implying that peak emission occurs for energies about five times larger than the Hawking temperature. In particular, a PBH with temperature $T_H\sim 1$ GeV emits more intensely particles with an energy $\sim 5$ GeV. Taking into account that the Hawking temperature is given by:
\begin{equation}
    T_{H}=   \frac{1  }{ (2+(\sqrt{1-\tilde{a}})^{-1})}{M_P^2\over 8\pi M}
\end{equation}
and that for the PBHs away from extremality ($\tilde{a}\lesssim 0.5$) that we are most interested in the first multiplicative factor is close to unity, we conclude that the use of PYTHIA is appropriate for PBH masses $\lesssim 10^{10}$ kg ($T_H\gtrsim 1$ GeV). We note that PYTHIA offers the possibility of extending its operative range via extrapolation tables. However, as described in \cite{Coogan:2020tuf}, this may lead to unreliable spectra. In particular, a comparison of the PYTHIA extrapolated spectra with the ones obtained using Hazma, for $M> 2.5 \times 10^{10}$ kg, reveals the failure of the former method in describing physical features such as neutral pion decay, $\pi^0\rightarrow \gamma\gamma$, which should yield a symmetric emission peaked at energies corresponding to half of the pion's mass that is not well reproduced by the PYTHIA-based spectra. The contribution of pion decay to the photon spectrum is significant for PBH masses in the range $2.5\times10^{10}-10^{11}$ kg ($T_H\sim 100-400$ MeV), since for lighter PBHs it is overcome by other secondary photon sources, while for heavier PBHs primary pion emission is Boltzmann-suppressed. 

For these reasons, we employ PYTHIA for PBH masses $M=5\times10^7-2.5\times10^{10}$ kg ({\it low mass range}), while for the interval $M=2.5\times 10^{10}-5\times10^{11}$ kg ({\it intermediate mass range}) we use Hazma, in both cases through the BlackHawk code. 

For PBH masses $M=5\times10^{11}-10^{12}$ kg ({\it high mass range}), for which the Hawking temperature $T_H\lesssim 20$ MeV, we have chosen to use the semi-analytical method described in  \cite{Coogan:2020tuf}, since secondary emission is in this case fully dominated by electron FSR, with no muons, pions, or QCD degrees of freedom being significantly emitted by the PBHs. In this case, the secondary photon spectrum is given by convoluting the primary electron and positron spectrum given in Eq.~(\ref{prim}) with the Altarelli-Parisi splitting functions at leading order in the electromagnetic fine-structure constant $\alpha_{EM}$ \cite{Altarelli:1977zs,Chen:2016wkt}:
\begin{equation}
{d^2N_{S, \gamma}\over dt dE_\gamma}= {\sum_{i=e^\pm, \mu^\pm, \pi^\pm}} \int dE_i {d^2N_{P, i}\over dt dE_i} {d^2N^{FSR}_{i}\over dE_\gamma},
\end{equation}
where for completeness we have also included the (sub-dominant) contributions from muons and charged pions and:
\begin{equation}
     {dN^{FSR}_{i}\over dE_\gamma}=\frac{\alpha_EM}{\pi Q_i}P_{i \rightarrow i\gamma}(x) \log \left(\frac{1-x}{\mu^2_i}-1\right),
\end{equation}
\begin{equation}
     P_{i \rightarrow i\gamma}(x)= \begin{cases}
  \frac{1+(1-x)^2}{x} & \text{for } i= e^\pm\text{, }\mu^\pm \\
  \frac{2(1-x)}{x} &  \text{for } i=\pi^\pm
  \end{cases},
\end{equation}
with $x= E_\gamma/E_i$, $\mu_i=m_i/2E_i$. With these analytical expressions for the splitting functions, we have computed the convolution integrals numerically for the {\it high mass range}, using the primary emission spectrum given in Eq.~(\ref{prim}) for the corresponding particles.

%%%%%%%%%%%%%%%%%%%%%%%%%%%%%%%%%%%%%%%%%%%%%%%%%%%%%%%%%%%%%%%%%%%%%%%%%%%
%%%%%%%%%%%%%%%%%%%%%%%%%%%%%%%%%%%%%%%%%%%%%%%%%%%%%%%%%%%%%%%%%%%%%%%%%%%
%%%%%%%%%%%%%%%%%%%%%%%%%%%%%%%%%%%%%%%%%%%%%%%%%%%%%%%%%%%%%%%%%%%%%%%%%%%
%%%%%%%%%%%%%%%%%%%%%%%%%%%%%%%%%%%%%%%%%%%%%%%%%%%%%%%%%%%%%%%%%%%%%%%%%%%

\section{Results for the Hawking spectrum}

In this section we show the results obtained through the methods described above for the full photon spectrum of PBHs for different values of their mass and spin, considering the three different mass ranges defined earlier. We will see that the spectrum exhibits different features in these different mass intervals, which will also determine employing distinct methodologies for mass and spin determination, as we discuss in the next section.

\subsection{Low mass range $(5\times10^7-2.5\times10^{10}\ \mathrm{kg})$}

\begin{figure}[htbp]
\centering\includegraphics[width=0.55\columnwidth]{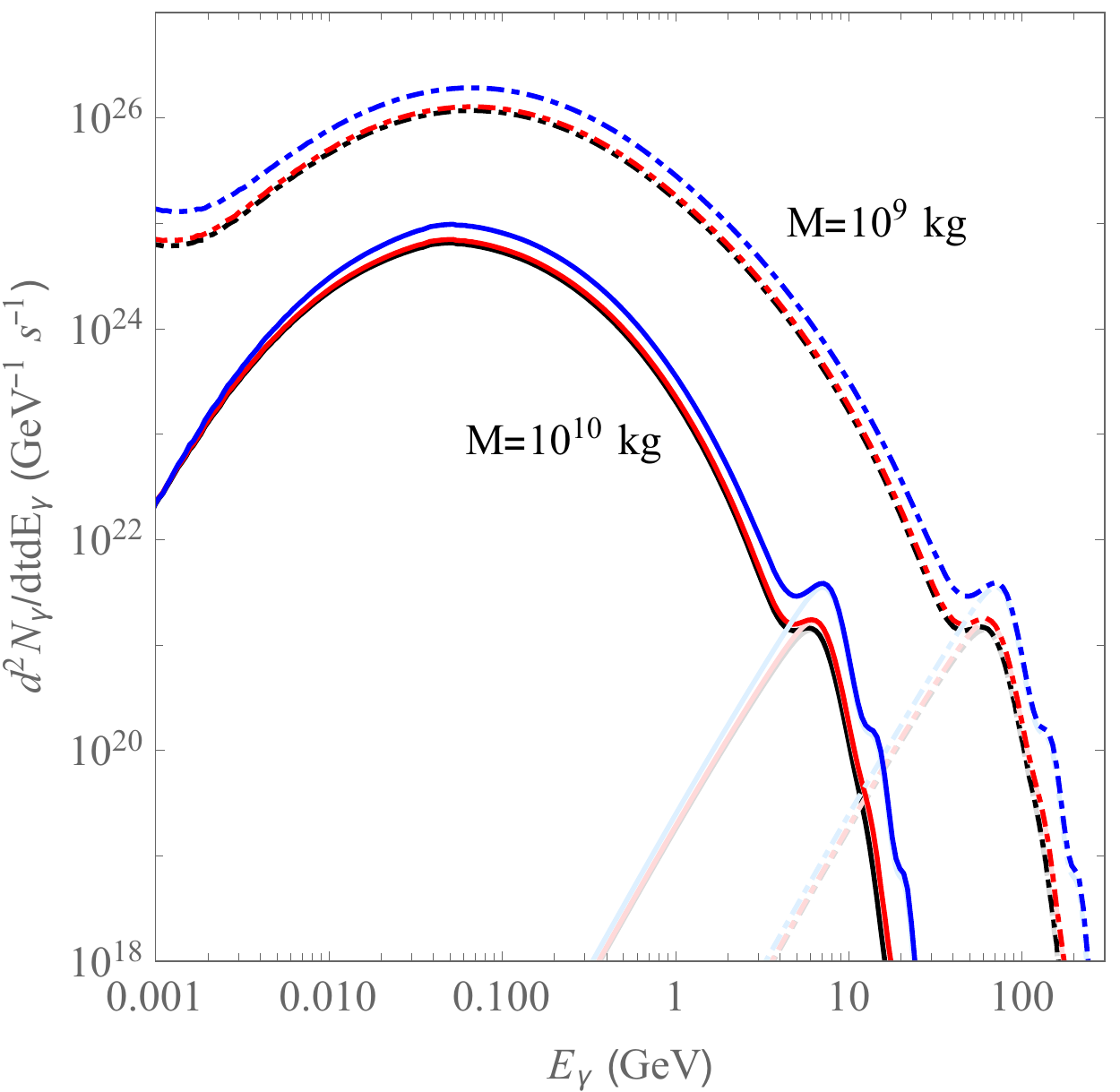}
\caption{Primary (light colors) and total (dark colors) photon emission rates for a PBH of $10^{10}$ kg (solid lines) and $10^9$ kg (dash-dotted lines), for  $\tilde{a}=0,0.2,0.5$ (black, red, and blue, respectively). The spectra are obtained using BlackHawk/PYTHIA.}\label{fig1}
\end{figure}

Fig.~\ref{fig1} shows the spectra obtained using BlackHawk and PYTHIA, for different PBH masses and spins. As one can clearly observe, the particle emission rate increases with the PBH spin $\tilde{a}$ for fixed mass. This is a generic feature of the emission of particles with non-zero spin like the photon and the many charged fermions contributing to the secondary spectrum in this low-mass/high-temperature range. We note that the opposite behaviour is observed for scalar emission, which decreases with $\tilde{a}$ (see e.g.~Fig.~2 in \cite{Arbey:2020yzj}).

\begin{figure}[htbp]
\centering\includegraphics[width=0.55\columnwidth]{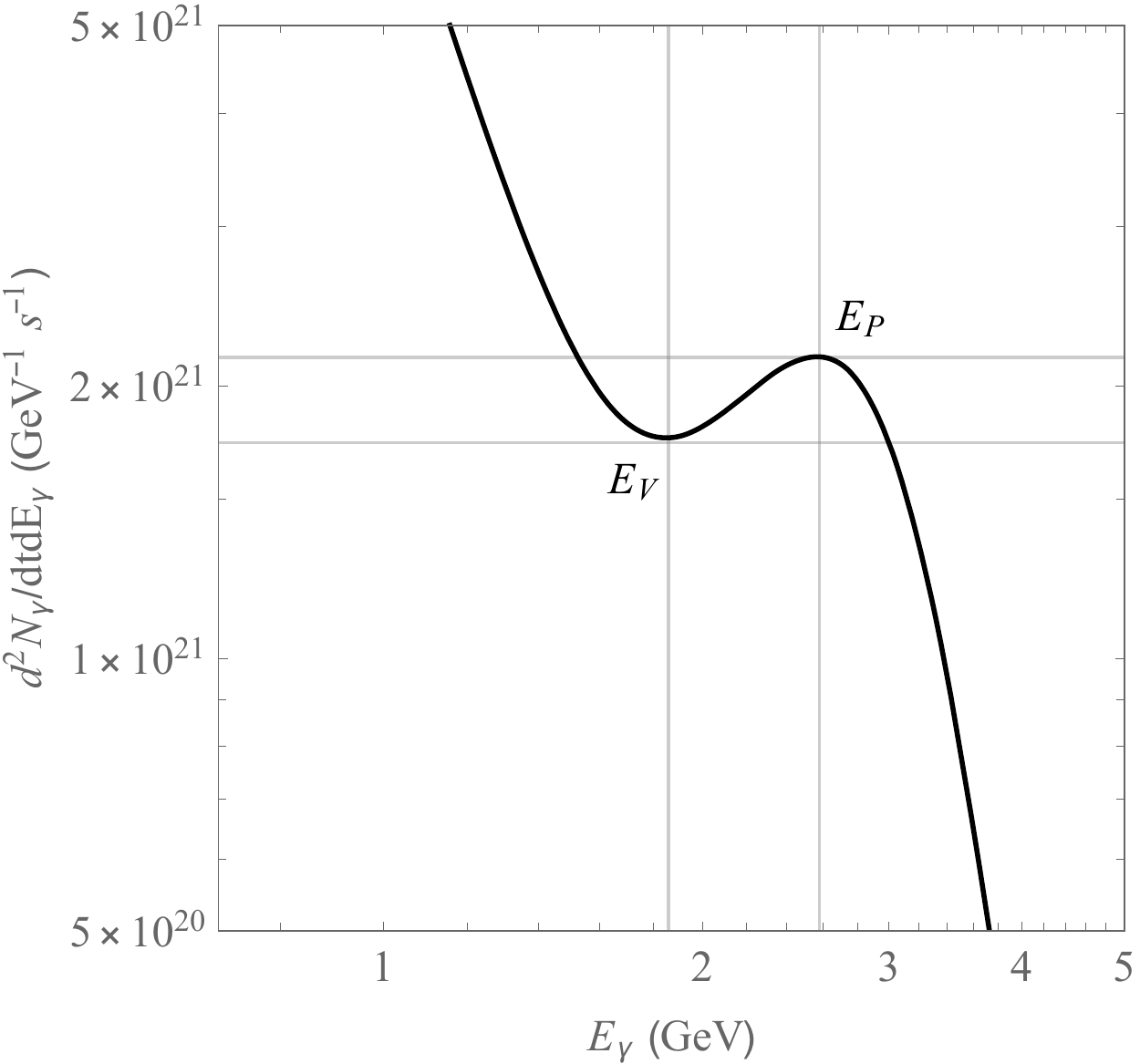}
\caption{Photon spectrum near the primary emission peak for $M=2.5\times10^{10}$ kg and $\tilde{a}=0.3$, illustrating the quantities characterizing the ``valley-peak'' structure.}
\label{fig2}
\end{figure}

In Fig.~\ref{fig1} we can also identify the primary emission peak at about five times the value of the Hawking temperature, while the secondary emission is more intense at lower energies. At the intersection of the secondary and primary emission spectra we can identify a ``valley-peak'' structure, and the energy of the ``valley", $E_V$, as identified in Fig.~\ref{fig2}, seems to be quite insensitive to the spin parameter $\tilde{a}$.

This is not surprising since the secondary spectrum depends on $\tilde{a}$ essentially through the primary spectrum of charged/unstable particles, which, like the primary photon spectrum, increases with $\tilde{a}$. Therefore, the energy at which they become comparable remains essentially the same for all PBH spins (for a given mass). The energy of the primary emission peak, $E_P$, is also largely independent of the PBH spin, due  to the mild dependence of the Hawking temperature on $\tilde{a}$ away from extremality, although there is a non-negligible spin dependence due to the graybody factors. In the next section, we will explore the mass and spin dependence of these two energy values to determine the latter values.

We note that the ``bell-shaped'' maximum of the secondary spectrum at lower energies, albeit more intense, cannot be used to reliably determine the PBH mass and spin, since its shape is degenerate in these parameters. For instance, we may increase the maximum emission rate and the broadness of the peak by either increasing $\tilde{a}$ or decreasing the PBH mass. Moreover, this bell structure is present in simulations performed using BlackHawk \cite{Arbey:2019mbc,Arbey:2021yke,Arbey:2021mbl} through the hadronization routine PYTHIA \cite{Sjostrand:2007gs,Bierlich:2022pfr}, but not when employing Hazma \cite{Coogan:2019qpu} or the semi-analytical methods proposed in \cite{MacGibbon:2015mya,Coogan:2020tuf}. For these reasons we will ignore this part of the secondary spectrum in our subsequent analysis.

\subsection{Intermediate mass range $(2.5\times10^{10}-5\times 10^{11}\ \mathrm{kg})$}

\begin{figure}[htbp]
\centering\includegraphics[width=0.55\columnwidth]{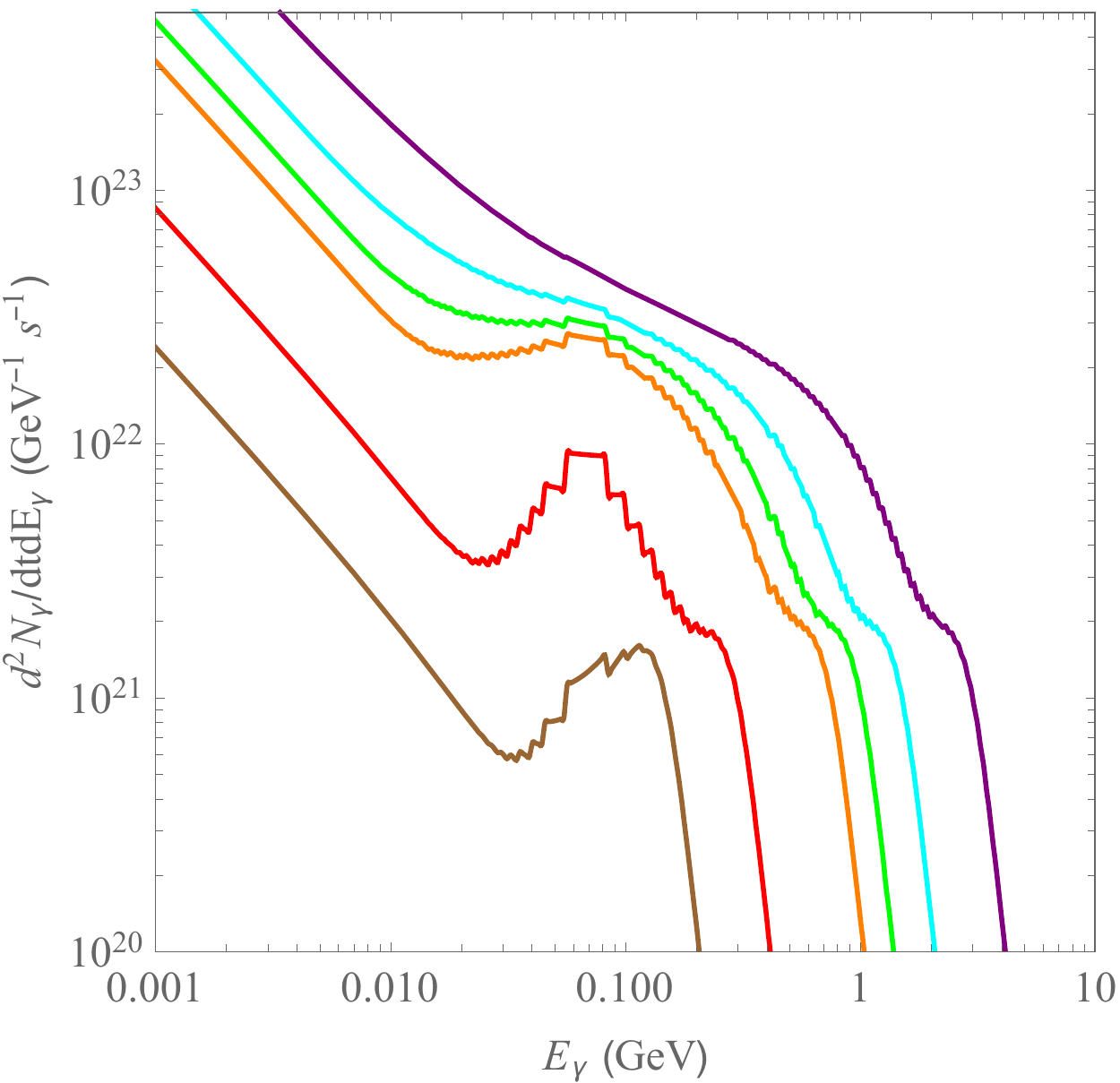}
\caption{Photon spectrum for $\tilde{a}=0$ PBHs with masses $2.5$, $5$, $7.5$, $10$, $25$ and $50 \times 10^{10}$ kg (purple, cyan, green, orange, red and brown, respectively), obtained using BlackHawk/Hazma.}\label{fig3}
\end{figure}

In this mass range we have used BlackHawk with Hazma to compute the secondary photon spectrum, particularly given the significant contribution of $\pi^0\rightarrow \gamma\gamma$, as visible in the spectra shown in Fig.~\ref{fig3} for $\tilde{a}=0$.

\begin{figure}[htbp]
\centering\includegraphics[width=0.57\columnwidth]{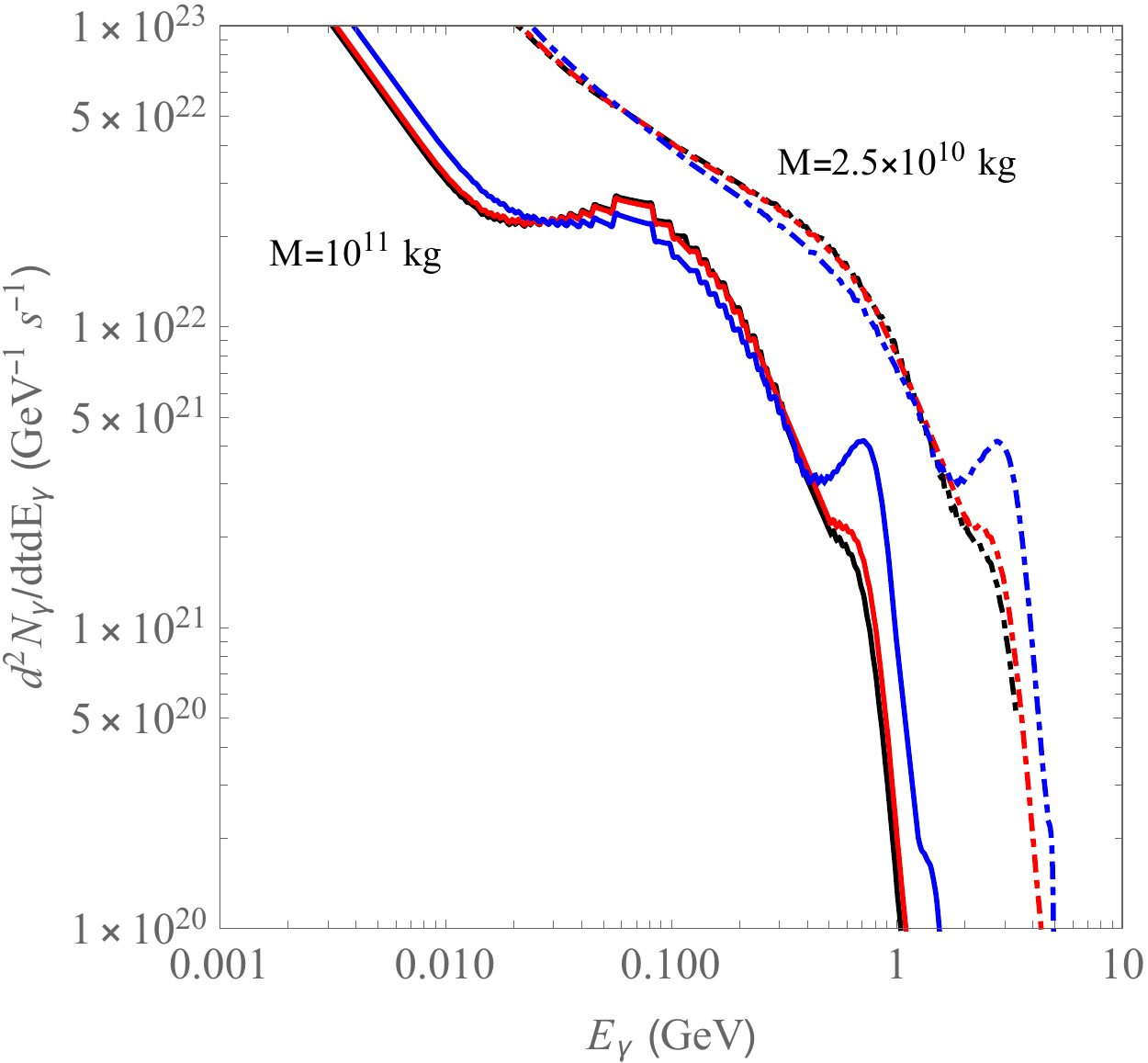}
\caption{Photon spectrum for PBHs with $\tilde{a}=0, 0.2, 0.5$ (black, red and blue, respectively) and for $M=2.5 \times 10^{10}$ and $10^{11}$ kg (dashed-dotted and solid lines, respectively), obtained using BlackHawk/Hazma.}\label{fig4}
\end{figure}

As one can see in this figure, pion decay induces a bump in the secondary spectrum at energies around half the pion mass ($\sim 0.07$ GeV). This feature is less prominent for the PBHs at the boundaries of this mass range since, on the one hand, for the lighter (hotter) ones other secondary emission processes overcome the pion contribution while, on the other hand, for the heavier (colder) ones primary pion emission is suppressed.
In Fig.~\ref{fig4} we show how the photon spectrum changes with the PBH spin, and as expected we see that the pion contribution decreases with $\tilde{a}$, given that they are spin-0 particles as explained earlier.

\begin{figure}[h]
\centering\includegraphics[width=0.57\columnwidth]{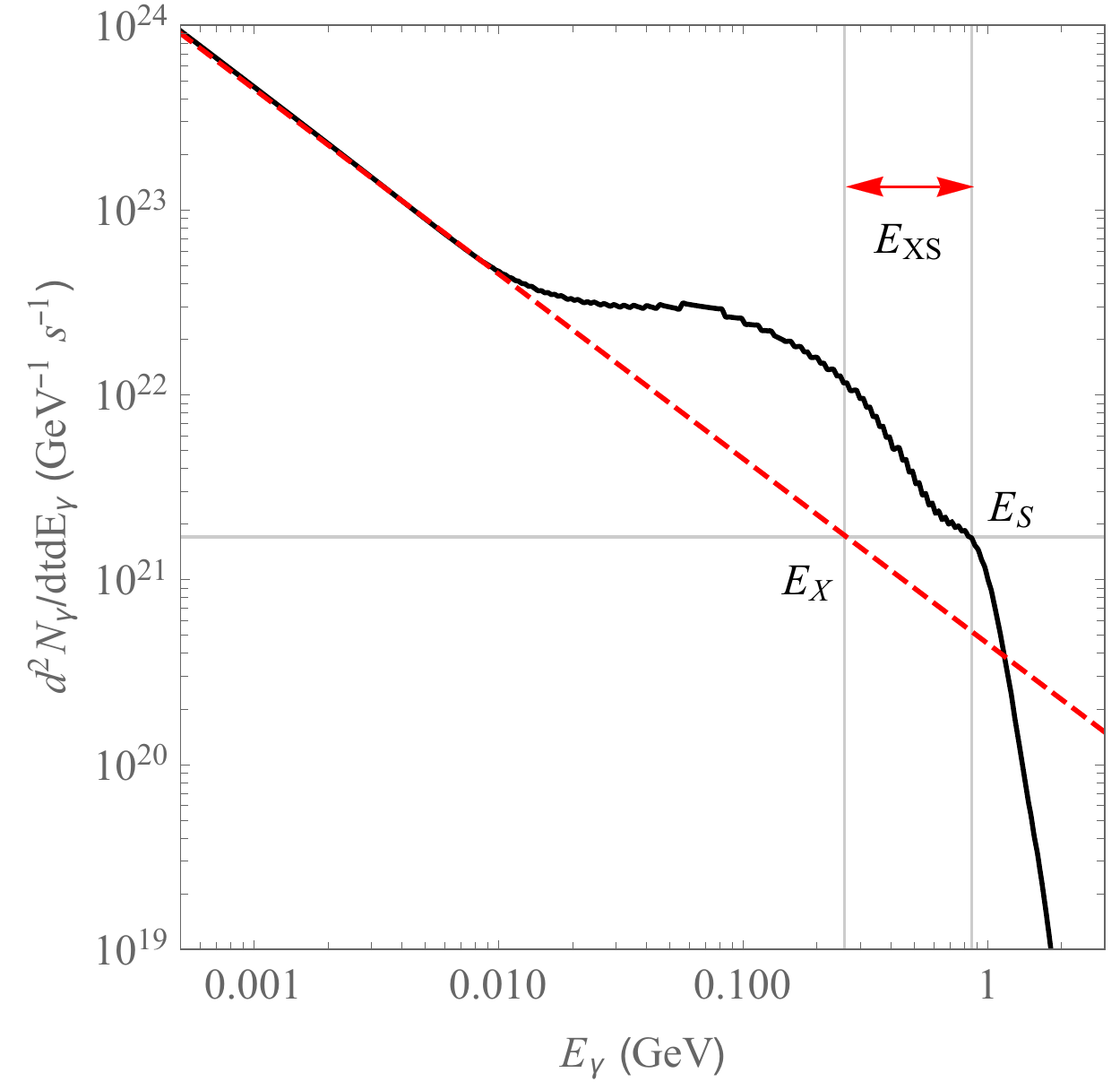}
\caption{Photon spectrum near the primary emission peak for a PBH with $M=7.5\times10^{10}$ kg and $\tilde{a}=0$, illustrating the ``shoulder'' feature, the pion ``bump'' and the fit to the low-energy tail of the secondary spectrum (dashed red line).}\label{fig5}
\end{figure}

A generic feature of the spectra in this mass range is, however, the absence of the ``valley-peak'' structure identified in the low mass range. Nevertheless, it is possible to characterize these spectra through alternative features that we will later use for mass and spin determination. In particular, we can identify a ``shoulder'' at the energies at which the primary and secondary photon emission rates become comparable, corresponding to a local maximum of $d^3N_\gamma/dtdE_\gamma^2$ at energy $E_S$, as shown in Fig.~\ref{fig5}. 

Another feature that we may use to characterize the spectrum is the energy $E_X$ at which the low-energy tail of the secondary spectrum (extrapolated  using a power-law fit) would match the emission rate at the ``shoulder'', in the absence of the pion ``bump'', as shown in Fig.~\ref{fig5}. We will examine the mass and spin dependence of these two energy values, as well as of their difference $E_{XS}=E_S-E_X$ in the next section.

\subsection{High mass range $(5\times10^{11}-10^{12}\ \mathrm{kg})$}

\begin{figure}[htbp]
\centering\includegraphics[width=0.54\columnwidth]{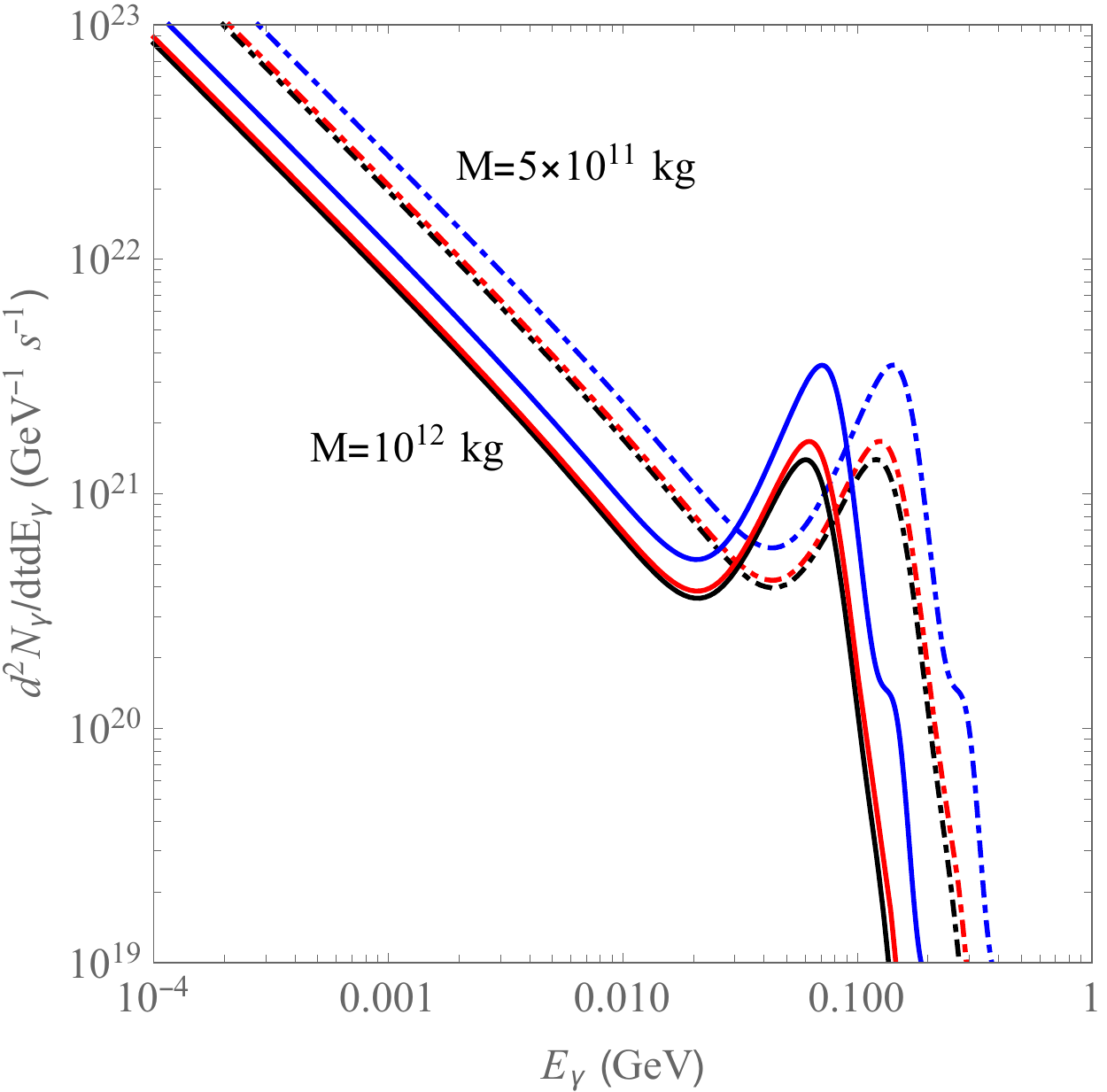}
\vspace{-0.3cm}
\caption{Photon spectrum for PBHs with $\tilde{a}=0, 0.2, 0.5$ (black, red and blue, respectively) and masses of $10^{11}$/$2.5 \times 10^{11}$ kg (solid/dashed-dotted lines), obtained using the semi-analytical method.}\label{fig6}
\end{figure}

\begin{figure}[H]
\centering\includegraphics[width=0.54\columnwidth]{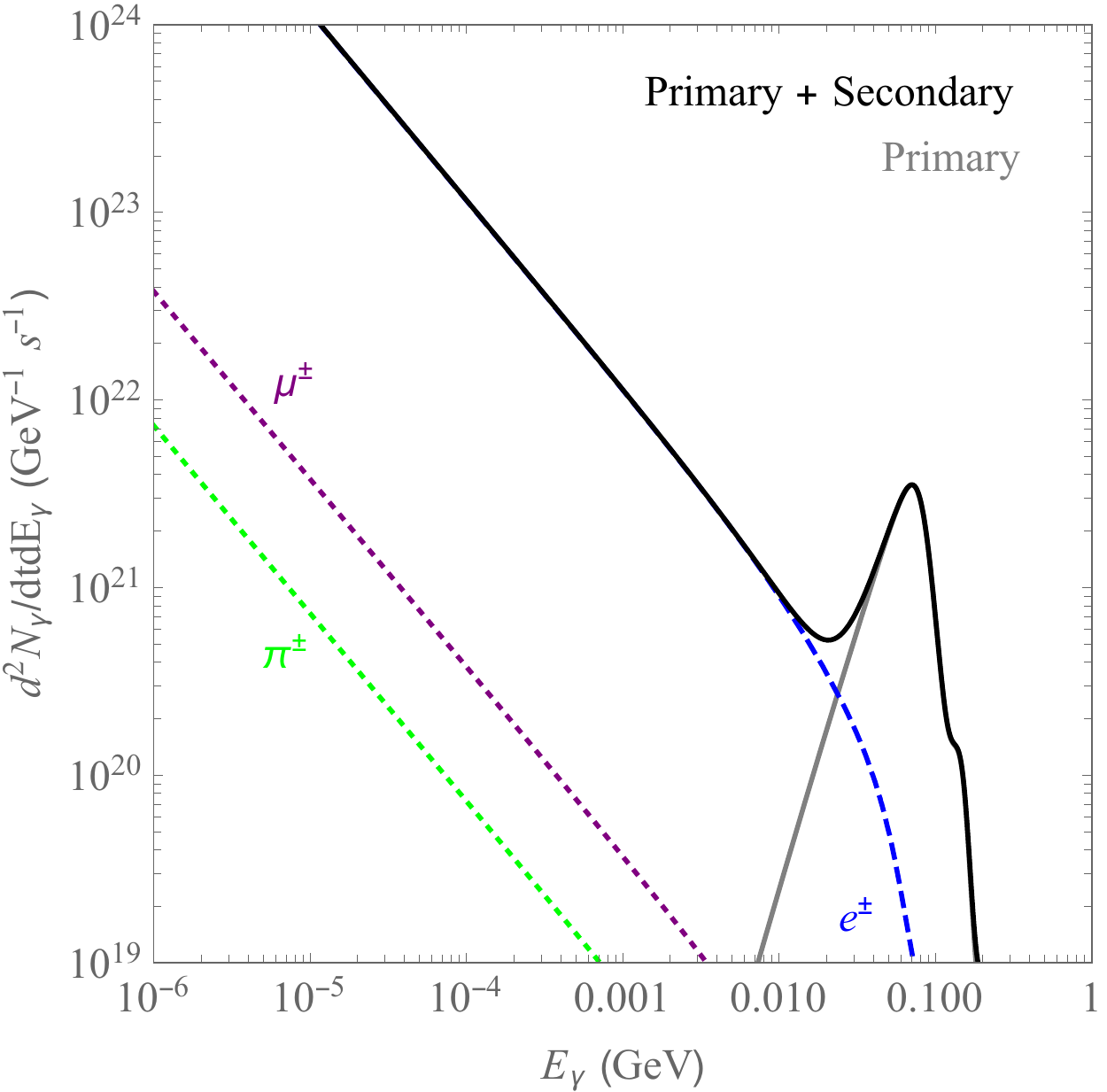}
\vspace{-0.3cm}
\caption{Photon spectrum for a Schwarzschild PBH with $M=5\times10^{11}$ kg, obtained using the semi-analytical method (black), illustrating the different contributions including primary photons (gray) and secondary photons from electron (blue), muon (purple) and pion (green) FSR.}\label{fig7}
\end{figure}

As described above, for these PBHs we computed the dominant FSR contributions to the secondary spectrum semi-analytically, our results being shown in Figs.~\ref{fig6} and \ref{fig7}. The pion ``bump'''s absence makes the ``valley-peak'' structure evident as for the low mass PBHs, and we may use the energies $E_P$ and $E_V$ to characterize the spectrum.

\vspace{-0.3cm}

Recall that PBHs with $M\gtrsim 10^{12}$ kg have not yet lost a significant fraction of their mass and hence could not have spun up due to axion emission. We nevertheless note that the spectrum is qualitatively similar for PBHs that can significantly emit electrons ($T_H\gtrsim m_e/5\simeq 0.1$ MeV or $M\lesssim 10^{14}$ kg), and is characterized by the same energy values.

%%%%%%%%%%%%%%%%%%%%%%%%%%%%%%%%%%%%%%%%%%%%%%%%%%%%%%%%%%%%%%%%%%%%%%%%%%%
%%%%%%%%%%%%%%%%%%%%%%%%%%%%%%%%%%%%%%%%%%%%%%%%%%%%%%%%%%%%%%%%%%%%%%%%%%%
%%%%%%%%%%%%%%%%%%%%%%%%%%%%%%%%%%%%%%%%%%%%%%%%%%%%%%%%%%%%%%%%%%%%%%%%%%%
%%%%%%%%%%%%%%%%%%%%%%%%%%%%%%%%%%%%%%%%%%%%%%%%%%%%%%%%%%%%%%%%%%%%%%%%%%%

\section{PBH mass and spin determination}

We now describe our proposal for determining the mass and spin of a PBH from its photon Hawking emission spectrum using the features identified in the previous section, within the different mass ranges of our analysis.

\subsection{Low mass range $(5\times10^7-2.5\times10^{10}\ \mathrm{kg})$}

\begin{figure}[htbp]
\centering\includegraphics[width=0.55\columnwidth]{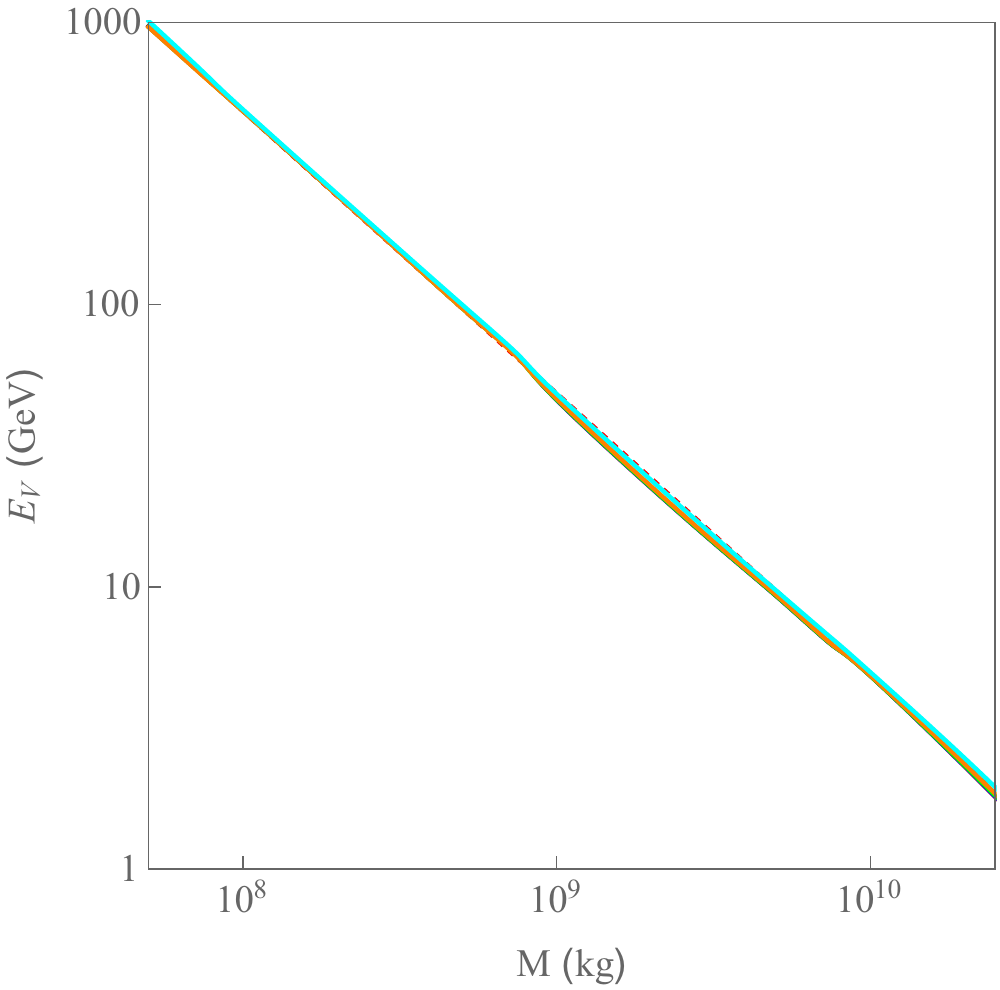}
\caption{Primary peak and ``valley'' energy as a function of the PBH mass for spin parameters $a_*=0, 0.1, 0.2, 0.3,0.5$ (red, blue, green, orange and cyan, respectively), computed using BlackHawk. The dotted red line yields the best fit curve for $E_V$ given by Eq.~(\ref{EV_fit_low}).}\label{fig8}
\end{figure}

In Fig.~\ref{fig8} we show how the energy $E_V$ depends on the PBH mass for different values of dimensionless spin parameter $\tilde{a}$. 

We thus find that the ``valley'' energy $E_V$ exhibits a simple dependence on the PBH mass, with only a very mild dependence on the spin parameter, which is well fitted by a simple inverse power law: 
\begin{equation} \label{EV_fit_low}
E_V^{(low)}\simeq {4.85 \times 10^{10}\ \mathrm {kg}\over M}\ \mathrm{GeV}~,
\end{equation}
such that a measurement of the ``valley'' energy can be used to determine the PBH mass with a relative error essentially given by the energy resolution of the detector, $\Delta E/E_V\simeq \Delta M/M$ (up to the mild dependence on $\tilde{a}$).

\begin{figure}[htbp]
\centering\includegraphics[width=0.56\columnwidth]{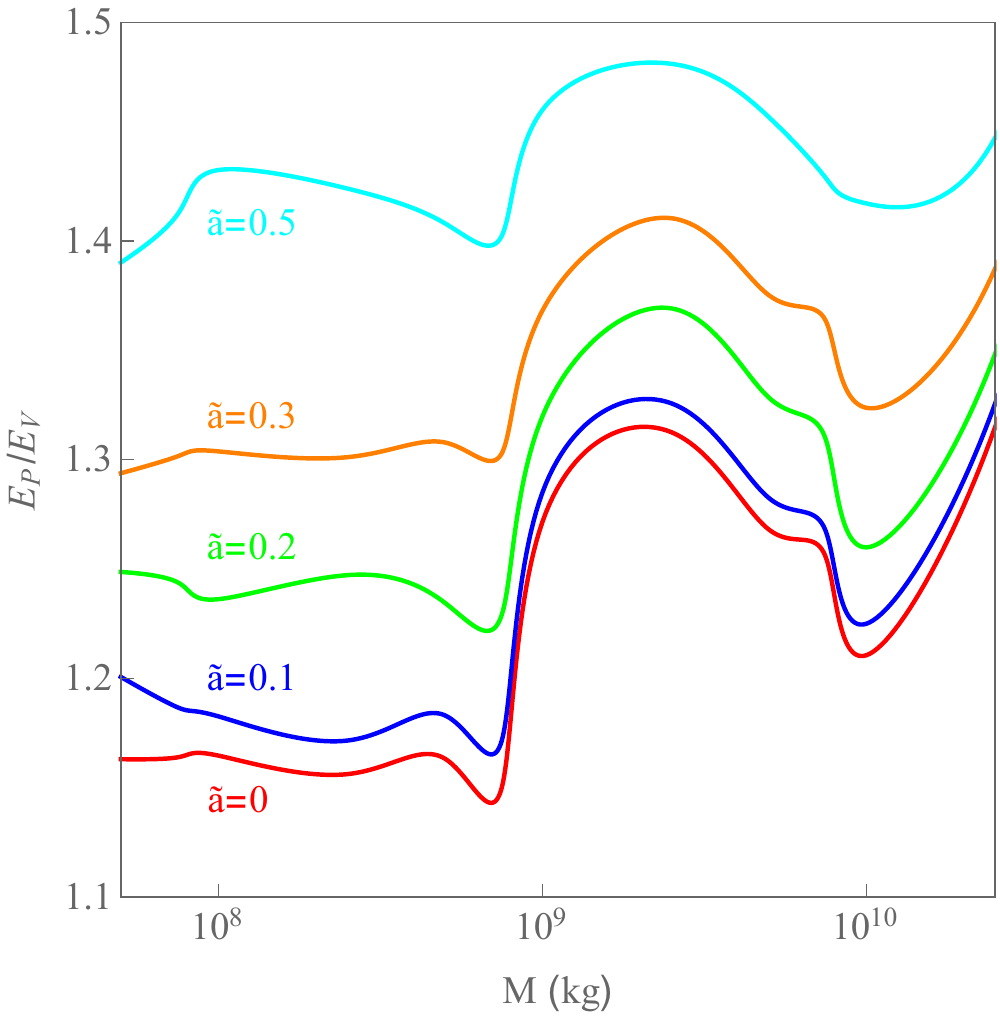}
\vspace{0.2cm}
\caption{Ratio $E_P/E_V$ as a function of the PBH mass computed with BlackHawk/PYTHIA, for $\tilde{a}=0, 0.1, 0.2, 0.3,0.5$ (red, blue, green, orange, cyan, respectively).}\label{fig9}
\end{figure}

On the other hand, the primary peak energy, albeit also quite close to a $M^{-1}$ dependence, exhibits a more pronounced dependence on the PBH spin, which is mostly due to the graybody factors, since as discussed earlier the Hawking temperature is only mildly dependent on this parameter (away from extremality as we are considering). Given the similar mass dependence of both $E_P$ and $E_V$, we expect their ratio to be largely independent of the PBH mass, therefore isolating the PBH spin dependence of the Hawking spectrum. This is illustrated in Fig.~\ref{fig9}, where we plot $E_P/E_V$ as a function of the PBH mass for different $\tilde{a}$ values.

\vspace{0.5cm}

This figure shows that the ratio $E_P/E_V$ has a non-trivial behaviour with the PBH mass, which essentially reflects the number of different SM particle species that are emitted by PBHs with different masses, particularly the effects of quarks and gluons.  Nevertheless, this ratio varies by less than 20\% for a fixed PBH spin within this mass range. This means that, if we determine the PBH mass from a measurement of $E_V$, a measurement of $E_P$ and of the ratio $E_P/E_V$ can be used to infer the PBH spin. Roughly, we conclude that measuring the energy of these features with a 10\% resolution may allow for distinguishing between a slowly rotating PBH ($\tilde{a}\ll 1$) and a PBH with $\tilde{a}\gtrsim 0.2$, i.e.~one that has spun up as it evaporated. If $\tilde{a}\lesssim 0.1$, a much better resolution below the few percent level should be required for inferring a non-vanishing rotation.

\subsection{Intermediate mass range $(2.5\times10^{10}-5\times 10^{11}\ \mathrm{kg})$}

\begin{figure}[htbp]
\centering\includegraphics[width=0.55\columnwidth]{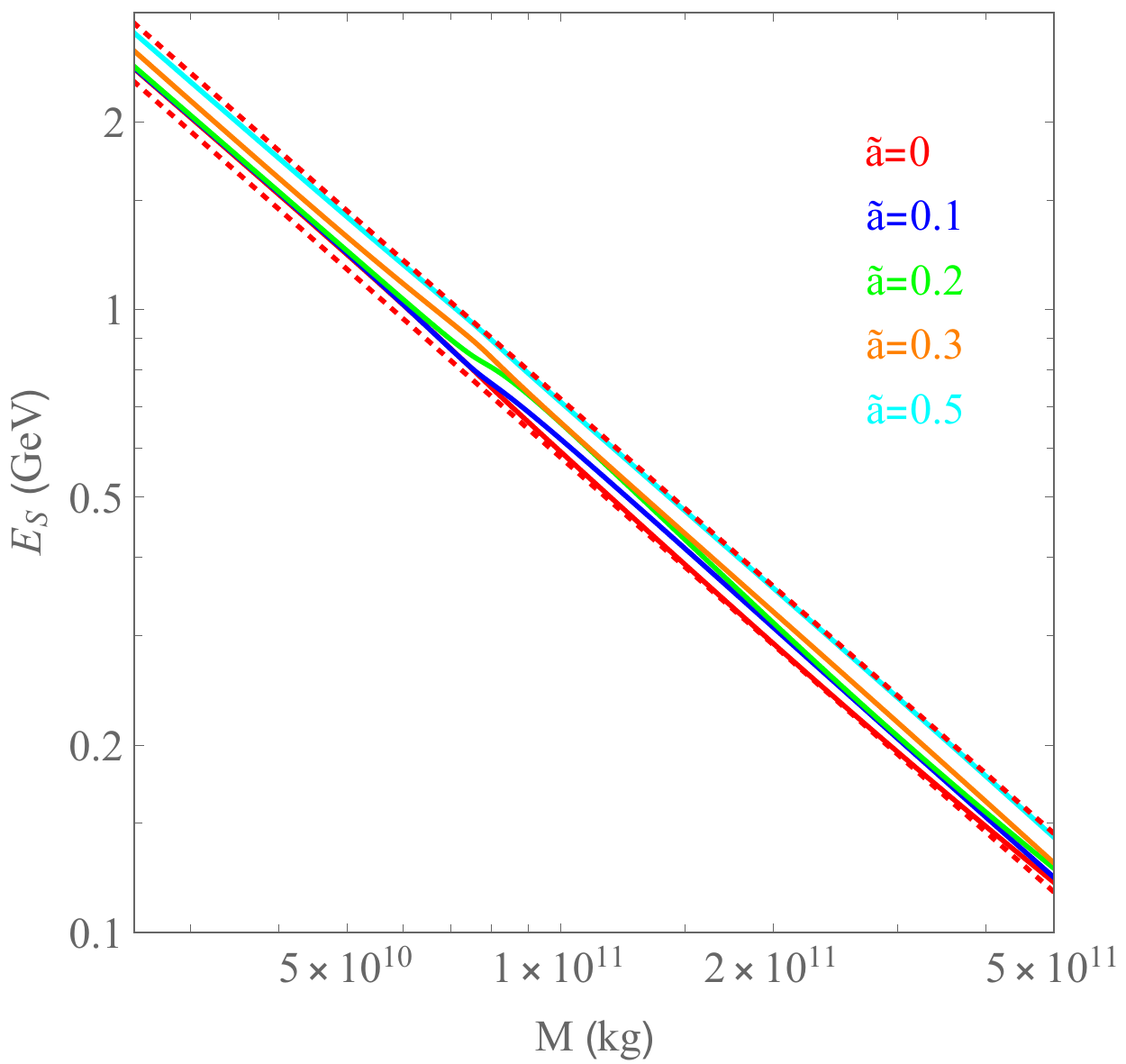}
\caption{``Shoulder'' energy $E_S$ as a function of the PBH mass computed using BlackHawk and Hazma, for $a_*=0, 0.1, 0.2, 0.3,0.5$ (red, blue, green, orange and cyan, respectively).}\label{fig10}
\end{figure}

In this mass range we will explore the mass and spin dependence of the energy features $E_S$ (``shoulder'' energy) and $E_{XS}=E_S-E_X$ identified in the previous section, given the absence of a clear ``valley-peak'' structure due to pion contamination. Unfortunately, none of these quantities exhibits a clean dependence on either the mass or the spin of PBH, so a precise determination of these quantities does not seem possible with this method. Nevertheless, as shown in Fig.~\ref{fig10}, $E_S$ is nearly inversely proportional to the PBH mass, although the proportionality constant is somewhat dependent on the PBH spin ($\lesssim 20\%$ variation for $\tilde{a}=0-0.5$). This means that a measurement of $E_S$ alone can at most allow us to infer the PBH mass with a $\lesssim $20\% error even with infinite energy resolution.

\begin{figure}
\centering\includegraphics[width=0.55\columnwidth]{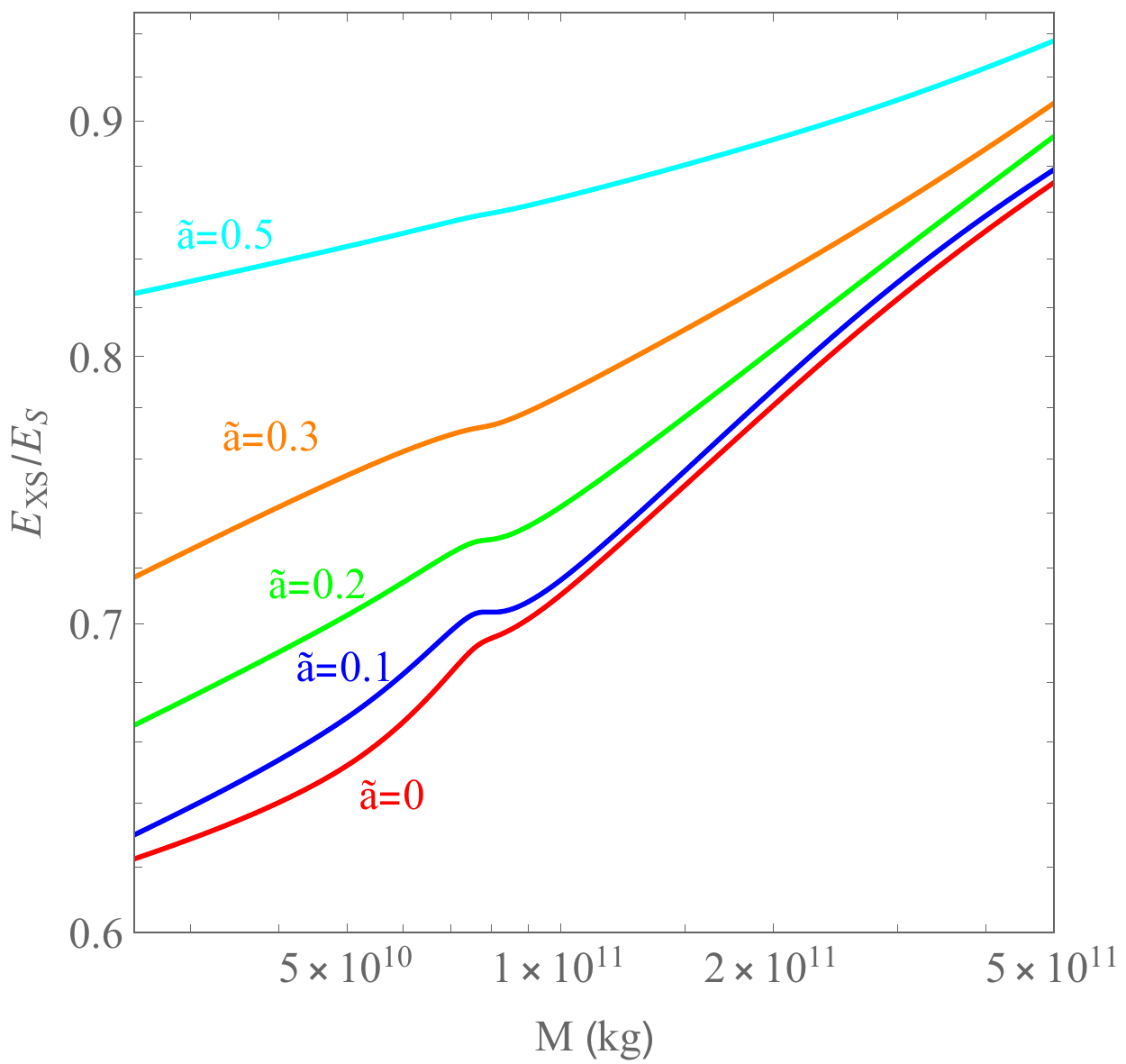}
\caption{Ratio $E_{XS}/E_S$ as a function of the PBH mass computed using BlackHawk/Hazma, for $a_*=0, 0.1, 0.2, 0.3,0.5$ (red, blue, green, orange and cyan, respectively).}\label{fig11}
\end{figure}

We now turn to the ratio $E_{XS}/E_S$, which is shown in Fig.~\ref{fig11} and exhibits essentially a power-law increase with the PBH mass within this range, despite some minor features. The exact power depends on $\tilde{a}$, making an accurate determination of the PBH spin challenging in this case, taking into account the uncertainty in measuring the PBH mass. In this case, we may nevertheless hope to obtain upper and lower bounds on both quantities depending on the energy resolution of the detector. For the lower PBH masses in this range, one would need to measure $E_{XS}/E_S$ with at least $\sim 5\%$ precision to infer $\tilde{a}\gtrsim 0.2$, while a percent level precision may be required for this towards the upper bound of this mass range.

We expect that these uncertainties might be mitigated by including additional features of the Hawking spectrum in the analysis, such as the height and width of the ``pion bump'', which are sensitive to both the PBH mass and spin as previously discussed (here we mean the height relative to e.g. the primary peak, such that this is independent of the PBH-Earth distance as well). While this may improve the precision with which the mass and spin of a PBH are determined, we will not pursue this any further in this work, given the theoretical uncertainties that could still plague the computation of the Hawking spectrum in this mass range, given in particular the discrepancies we have observed in the spectra obtained using PYTHIA or Hazma alongside BlackHawk. The method proposed for this mass range relies mostly on the primary photon emission (which determines the ``shoulder'' position to a large extent) and the low-energy tail of the secondary emission, so while it cannot clearly disentangle the mass and spin dependence of the spectrum it is nevertheless robust from the theoretical perspective, and a more thorough analysis of other features is left for future work.

\subsection{High mass range $(5\times10^{11}-10^{12}\ \mathrm{kg})$}

\begin{figure}[htbp]
\centering\includegraphics[width=0.55\columnwidth]{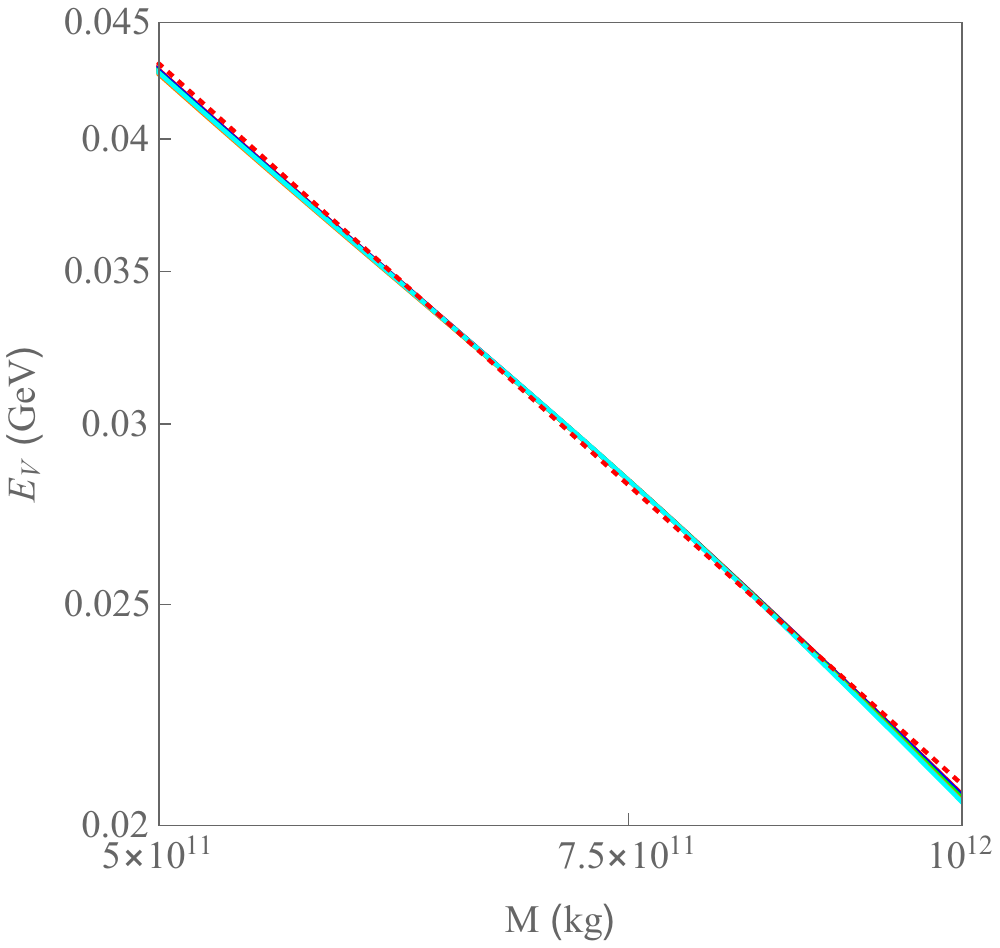}
\caption{``Valley'' energy as a function of the PBH mass, computed with the semi-analytical method, for $\tilde{a}=0, 0.1, 0.2, 0.3,0.5$ (red, blue, green, orange and cyan, respectively). The dotted red line yields the best fit curve in Eq.~(\ref{EV_fit_high}).}\label{fig12}
\end{figure}

For masses higher then $5 \times 10^{11}$ kg, the pion contribution becomes subdominant and the ``valley-peak'' structure is again well defined, so that we may use the method employed for the low mass range. In Fig.~\ref{fig12} we show how $E_V$ and $E_P$ depend on the PBH mass for different values of $\tilde{a}$ in this mass range.

\begin{figure}[H]
\centering\includegraphics[width=0.55\columnwidth]{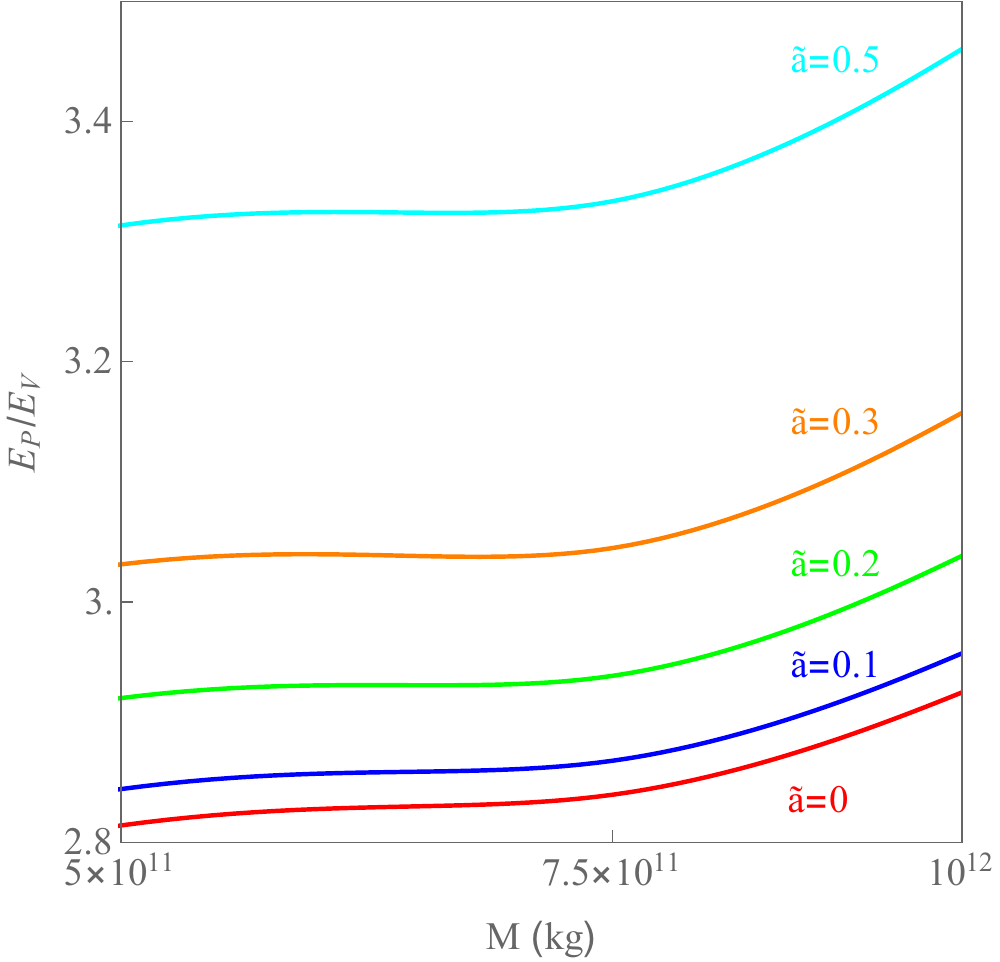}
\caption{Ratio $E_P/E_V$ as a function of the PBH mass, computed using the semi-analytical method, for $\tilde{a}=0, 0.1, 0.2, 0.3,0.5$ (red, blue, green, orange and cyan, respectively).}\label{fig13}
\end{figure}

As in the low mass range, $E_V$ is nearly independent of the PBH spin, although its mass dependence deviates slightly from an inverse power law: 
\begin{equation} \label{EV_fit_high}
E_V^{(high)}\simeq \left({2.5 \times 10^{10}\ \mathrm {kg}\over M}\right)^{1.05}\ \mathrm{GeV}~,
\end{equation}
In Fig.~\ref{fig13} we plot the ratio $E_P/E_V$ as a function of the PBH mass for different spin values, and in this mass range we find only a mild dependence on the mass, such that measuring this ratio with a precision of a few percent suffices to distinguish a non-spinning PBH from one that is spinning with $\tilde{a}\gtrsim 0.2$, with around 10\% resolution required to measure $\tilde{a}\gtrsim 0.5$.

%%%%%%%%%%%%%%%%%%%%%%%%%%%%%%%%%%%%%%%%%%%%%%%%%%%%%%%%%%%%%%%%%%%%%%%%%%%
%%%%%%%%%%%%%%%%%%%%%%%%%%%%%%%%%%%%%%%%%%%%%%%%%%%%%%%%%%%%%%%%%%%%%%%%%%%
%%%%%%%%%%%%%%%%%%%%%%%%%%%%%%%%%%%%%%%%%%%%%%%%%%%%%%%%%%%%%%%%%%%%%%%%%%%
%%%%%%%%%%%%%%%%%%%%%%%%%%%%%%%%%%%%%%%%%%%%%%%%%%%%%%%%%%%%%%%%%%%%%%%%%%%

\newpage
\section{Conclusion}

In this work we have proposed a distance-independent methodology for determining the mass and spin of a PBH from its photon Hawking emission spectrum, including both the primary and secondary emission components. The methods rely on the energy of particular features in the spectrum, in the vicinity of the primary emission peak. For most of the PBH mass range considered ($5\times10^7-10^{12}$ kg), corresponding to PBHs that have already lost a significant fraction of their mass through evaporation, these are the primary peak energy and the ``valley'' energy, the latter corresponding to the point where the secondary and primary emission spectra become comparable. Using both these energy values, one can partially disentangle the effects of the PBH mass and spin, allowing for an accurate measurement of both quantities with a detector with sufficient energy resolution. We generically find that an energy resolution of at least a few percent may allow for distinguishing a non-spinning PBH ($\tilde{a}\ll 1$) from one that is spinning with $\tilde{a}\gtrsim 0.2$. This methodology fails only in the intermediate PBH mass range ($2.5\times10^{10}-5\times10^{11}$ kg) where neutral pion decays significantly change the shape of the spectrum, but even in this case we were able to identify spectral features that at least allow one to place lower and upper bounds on the PBH mass and spin.

As the proposed methods rely on energy values rather than photon emission rates, they are independent of the Earth-PBH distance, $d$. Once $M$ and $\tilde{a}$ are known, one can then predict the expected photon flux by dividing the theoretical emission rate $d^2N_\gamma/dtdE_\gamma$ by $4\pi d^2$, such that the observed photon counts can be used to infer the distance $d$. If the PBH is close enough for $d$ to also be accurately determined through parallax measurements, this can be used to check the validity of the evaporating PBH hypothesis and also the consistency of the mass and spin determination.

We should note, however, that it may be challenging to, in practice, implement our proposed methodology, since it relies on the shape of the spectrum close to the primary emission peak, where emission rates are considerably lower than for the low-energy tail of the secondary spectrum (see e.g. Fig.~\ref{fig1}). If PBHs in this mass range do exist, they are much more likely to be found through the latter part of the spectrum, and only after detector sensitivities can be improved by a few orders of magnitude will it be possible to study spectral features close to the primary emission peak. For example, the proposed AMEGO-X \cite{AMEGO:2019gny, Fleischhack:2021mhc} and All-Sky-ASTROGAM missions \cite{e-ASTROGAM:2016bph, Tatischeff:2019mun} are expected to have sensitivities down to $\sim 10^{-6}\ \mathrm{MeV}\mathrm{cm}^{-2}\mathrm{s}^{-1}$ for photon energies $E_\gamma\sim 1\ \mathrm{MeV} - 1\ \mathrm{GeV}$ (and energy resolution $<10$\%), which can be recast as a lower bound on the PBH emission rate
\begin{equation} \label{AMEGO}
{d^2N_\gamma\over dtdE_\gamma} \gtrsim 10^{24}\!\left({0.1\ \mathrm{GeV}\over E_\gamma}\right)^{\!2}\!\left({d\over 100\ \mathrm{AU}}\right)^{\!2}\mathrm{GeV}^{-1}\mathrm{s}^{-1}~.
\end{equation}
This is above the primary peak emission rate for the PBH mass range considered except for PBHs within only a few AU of the Earth, which we would have to be extremely lucky to detect. Note also that only for the heaviest PBHs with $E_P\lesssim 1$ GeV could the primary spectrum be within the reach of such detectors. The ongoing Fermi-LAT instrument has a comparable sensitivity at higher energies, and PBHs searches with this instrument have already been performed, albeit so far with null results \cite{Fermi-LAT:2018pfs}. The proposed MAST mission \cite{Dzhatdoev:2019kay}, with a large effective area, could improve upon Fermi-LAT's state-of-the-art sensitivity for high-energy gamma-rays (100 MeV - 1 TeV) by one order of magnitude, with an energy resolution possibly down to 6\%-10\% above 10 GeV. This may potentially enable the detection of primary Hawking emission for PBHs with $M\lesssim 10^{10}$ kg and the implementation of our mass and spin determination methods. PBH detection in the neutrino channel has also been proposed in the literature \cite{Halzen:1995hu, Dave:2019epr, Capanema:2021hnm, Bernal:2022swt}, and we envisage that the neutrino Hawking spectrum may, in principle, provide additional information for mass and spin determination, which we plan to investigate in future work.

Our results show that it may be possible in the near future to not only detect Hawking radiation from small PBHs but also determine whether or not they rotate with moderately large spins, $\tilde{a}\gtrsim 0.2$. Evidence for the existence of such spinning PBHs could have a considerable impact in high-energy physics, given that PBHs lighter than $10^{12}$ kg should already have lost essentially all their angular momentum through Hawking emission of Standard Model particles and, to our knowledge, only the emission of a large number of light scalars (characteristic of scenarios such as the string axiverse) could justify finding $\tilde{a}\sim 0.1-0.5$. We thus hope that this work serves as further motivation for planning future high-energy gamma-ray telescopes with better sensitivity.

%We provided a brief discussion of the nowadays available numerical and semi-analytical tools for calculating a BH spectrum and discussed their limits of validity.
%Afterwards, we presented and discussed the main feature of BH spectra obtain with those tools with in their validity ranges.
%We outlined how the particle physics influence the spectrum of a BH, and the necessity to adopt different mass and spin extrapolation methods according to the energy range and underling particle physics processing occurring. Two different methods have been presented and discussed.
%Finally we showed how the information gained in this analysis can be used in order to infer the distance of a BH. 

\begin{acknowledgments}
M.C. is supported by the FCT doctoral grant SFRH/BD/146700/2019. This work was supported by national funds from FCT - Funda\c{c}\~ao para a Ci\^encia e a Tecnologia, I.P., within the project UID/04564/2020 and the grant No.~CERN/FIS-PAR/0027/2021.
\end{acknowledgments}


\begin{thebibliography}{99}

%\cite{Hawking:1971ei}
\bibitem{Hawking:1971ei} 
  S.~Hawking,
  %``Gravitationally collapsed objects of very low mass,''
  Mon.\ Not.\ Roy.\ Astron.\ Soc.\  {\bf 152}, 75 (1971).
  %%CITATION = MNRAA,152,75;%%
  %561 citations counted in INSPIRE as of 25 Nov 2019

%\cite{Carr:1974nx}
\bibitem{Carr:1974nx} 
  B.~J.~Carr and S.~W.~Hawking,
  %``Black holes in the early Universe,''
  Mon.\ Not.\ Roy.\ Astron.\ Soc.\  {\bf 168}, 399 (1974).
  %%CITATION = MNRAA,168,399;%%
  %413 citations counted in INSPIRE as of 27 Feb 2017
  
%\cite{Carr:1975qj}
\bibitem{Carr:1975qj} 
  B.~J.~Carr,
  %``The Primordial black hole mass spectrum,''
  Astrophys.\ J.\  {\bf 201}, 1 (1975).
  %%doi:10.1086/153853
  %%CITATION = %doi:10.1086/153853;%%
  %398 citations counted in INSPIRE as of 27 Feb 2017  
  
  
%\cite{Carr:2020xqk}
\bibitem{Carr:2020xqk}
B.~Carr and F.~Kuhnel,
%``Primordial Black Holes as Dark Matter: Recent Developments,''
Ann. Rev. Nucl. Part. Sci. \textbf{70}, 355-394 (2020).
%%doi:10.1146/annurev-nucl-050520-125911
%[arXiv:2006.02838 [astro-ph.CO]].
%149 citations counted in INSPIRE as of 23 Aug 2021

%\cite{Clesse:2017bsw}
\bibitem{Clesse:2017bsw}
S.~Clesse and J.~Garc\'\i{}a-Bellido,
%``Seven Hints for Primordial Black Hole Dark Matter,''
Phys. Dark Univ. \textbf{22}, 137-146 (2018).
%%doi:10.1016/j.dark.2018.08.004
%[arXiv:1711.10458 [astro-ph.CO]].
%146 citations counted in INSPIRE as of 24 May 2022

%\cite{Sasaki:2016jop}
\bibitem{Sasaki:2016jop}
M.~Sasaki, T.~Suyama, T.~Tanaka and S.~Yokoyama,
%``Primordial Black Hole Scenario for the Gravitational-Wave Event GW150914,''
Phys. Rev. Lett. \textbf{117}, no.6, 061101 (2016)
[erratum: Phys. Rev. Lett. \textbf{121}, no.5, 059901 (2018)].
%%doi:10.1103/PhysRevLett.117.061101
%[arXiv:1603.08338 [astro-ph.CO]].
%511 citations counted in INSPIRE as of 23 Aug 2021

%\cite{Niikura:2019kqi}
\bibitem{Niikura:2019kqi}
H.~Niikura, M.~Takada, S.~Yokoyama, T.~Sumi and S.~Masaki,
%``Constraints on Earth-mass primordial black holes from OGLE 5-year microlensing events,''
Phys. Rev. D \textbf{99}, no.8, 083503 (2019).
%%doi:10.1103/PhysRevD.99.083503
%[arXiv:1901.07120 [astro-ph.CO]].
%155 citations counted in INSPIRE as of 24 May 2022

\bibitem{PartCrea}
S.W.~Hawking, ``Particle creation by black holes,'' Comm. Math. Phys. 43
(1975) 199.

 %\cite{Carr:2009jm, Carr:2016hva, Carr:2020gox, Arbey:2019vqx}
\bibitem{Carr:2009jm} 
  B.~J.~Carr, K.~Kohri, Y.~Sendouda and J.~Yokoyama,
  %``New cosmological constraints on primordial black holes,''
  Phys.\ Rev.\ D{\bf 81}, 104019 (2010).
  %%doi:10.1103/PhysRevD.81.104019
  %[arXiv:0912.5297 [astro-ph.CO]].
  %%CITATION = %%doi:10.1103/PhysRevD.81.104019;%%
  %510 citations counted in INSPIRE as of 25 Nov 2019
  
  %\cite{Carr:2016hva}
\bibitem{Carr:2016hva}
B.~J.~Carr, K.~Kohri, Y.~Sendouda and J.~Yokoyama,
%``Constraints on primordial black holes from the Galactic gamma-ray background,''
Phys. Rev. D \textbf{94}, no.4, 044029 (2016)
%doi:10.1103/PhysRevD.94.044029
[arXiv:1604.05349 [astro-ph.CO]].
%107 citations counted in INSPIRE as of 30 Sep 2022

%\cite{Carr:2020gox}
\bibitem{Carr:2020gox}
B.~Carr, K.~Kohri, Y.~Sendouda and J.~Yokoyama,
%``Constraints on primordial black holes,''
Rept. Prog. Phys. \textbf{84}, no.11, 116902 (2021)
%doi:10.1088/1361-6633/ac1e31
[arXiv:2002.12778 [astro-ph.CO]].
%494 citations counted in INSPIRE as of 30 Sep 2022

%\cite{Arbey:2019vqx}
\bibitem{Arbey:2019vqx}
 A.~Arbey, J.~Auffinger and J.~Silk,
 %``Constraining primordial black hole masses with the isotropic gamma ray background,''
 Phys.\ Rev.\ D{\bf 101}, no. 2, 023010 (2020).
 %%doi:10.1103/PhysRevD.101.023010
 %[arXiv:1906.04750 [astro-ph.CO]].
 %%CITATION = %doi:10.1103/PhysRevD.101.023010;%%
 %8 citations counted in INSPIRE as of 06 Mar 2020 

%\cite{Ferraz:2020zgi}
\bibitem{Ferraz:2020zgi}
P.~B.~Ferraz, T.~W.~Kephart and J.~G.~Rosa,
%``Superradiant pion clouds around primordial black holes,''
JCAP \textbf{07}, no.07, 026 (2022)
%doi:10.1088/1475-7516/2022/07/026
[arXiv:2004.11303 [gr-qc]].
%5 citations counted in INSPIRE as of 30 Sep 2022

\bibitem{Auffinger:2022khh}
J.~Auffinger,
%``Primordial black hole constraints with Hawking radiation -- a review,''
[arXiv:2206.02672 [astro-ph.CO]].
%0 citations counted in INSPIRE as of 11 Jul 2022

\bibitem{foot1}
Note that 21cm signal observations made by the EDGES collaboration yield stronger constraints, although these depend on how the intergalactic medium is heated \cite{Mittal:2021egv}.

 
%\cite{Mittal:2021egv}
\bibitem{Mittal:2021egv}
S.~Mittal, A.~Ray, G.~Kulkarni and B.~Dasgupta,
%``Constraining primordial black holes as dark matter using the global 21-cm signal with X-ray heating and excess radio background,''
JCAP \textbf{03}, 030 (2022)
%doi:10.1088/1475-7516/2022/03/030
[arXiv:2107.02190 [astro-ph.CO]].
%28 citations counted in INSPIRE as of 07 Oct 2022

%\cite{Glicenstein:2013vha, Tavernier:2019exh}
\bibitem{Glicenstein:2013vha}
J.~F.~Glicenstein \textit{et al.} [H.E.S.S.],
%``Limits on Primordial Black Hole evaporation with the H.E.S.S. array of Cherenkov telescopes,''
[arXiv:1307.4898 [astro-ph.HE]].
%19 citations counted in INSPIRE as of 07 Oct 2022

%\cite{Tavernier:2019exh}
\bibitem{Tavernier:2019exh}
T.~Tavernier, J.~F.~Glicenstein and F.~Brun,
%``Search for Primordial Black Hole evaporations with H.E.S.S,''
PoS \textbf{ICRC2019}, 804 (2020)
%doi:10.22323/1.358.0804
[arXiv:1909.01620 [astro-ph.HE]].
%7 citations counted in INSPIRE as of 07 Oct 2022

%\cite{HAWC:2019wla}
\bibitem{HAWC:2019wla}
A.~Albert \textit{et al.} [HAWC],
%``Constraining the Local Burst Rate Density of Primordial Black Holes with HAWC,''
JCAP \textbf{04}, 026 (2020)
%doi:10.1088/1475-7516/2020/04/026
[arXiv:1911.04356 [astro-ph.HE]].
%15 citations counted in INSPIRE as of 07 Oct 2022

%\cite{Abdo:2014apa}
\bibitem{Abdo:2014apa}
A.~A.~Abdo, %A.~U.~Abeysekara, R.~Alfaro, B.~T.~Allen, C.~Alvarez, J.~D.~\'Alvarez, R.~Arceo, J.~C.~Arteaga-Vel\'azquez, T.~Aune and H.~A.~Ayala Solares, 
\textit{et al.}
%``Milagro Limits and HAWC Sensitivity for the Rate-Density of Evaporating Primordial Black Holes,''
Astropart. Phys. \textbf{64}, 4-12 (2015)
%doi:10.1016/j.astropartphys.2014.10.007
[arXiv:1407.1686 [astro-ph.HE]].
%33 citations counted in INSPIRE as of 07 Oct 2022

%\cite{Archambault:2017asc}
\bibitem{Archambault:2017asc}
S.~Archambault [VERITAS],
%``Search for Primordial Black Hole Evaporation with VERITAS,''
PoS \textbf{ICRC2017}, 691 (2018)
%doi:10.22323/1.301.0691
[arXiv:1709.00307 [astro-ph.HE]].
%14 citations counted in INSPIRE as of 07 Oct 2022

%\cite{Fermi-LAT:2018pfs}
\bibitem{Fermi-LAT:2018pfs}
M.~Ackermann \textit{et al.} [Fermi-LAT],
%``Search for Gamma-Ray Emission from Local Primordial Black Holes with the Fermi Large Area Telescope,''
Astrophys. J. \textbf{857}, no.1, 49 (2018).
%%doi:10.3847/1538-4357/aaac7b
%[arXiv:1802.00100 [astro-ph.HE]].
%14 citations counted in INSPIRE as of 23 Aug 2021


%\cite{}
\bibitem{Halzen:1990ip}
F.~Halzen, E.~Zas, J.~H.~MacGibbon and T.~C.~Weekes,
%``Search for gamma-rays from black holes,''
MAD-PH-575.
%0 citations counted in INSPIRE as of 28 Apr 2022

%\cite{}
\bibitem{Halzen:1991uw}
F.~Halzen, E.~Zas, J.~H.~MacGibbon and T.~C.~Weekes,
%``Gamma-rays and energetic particles from primordial black holes,''
Nature \textbf{353}, 807-815 (1991).
%%doi:10.1038/353807a0
%140 citations counted in INSPIRE as of 28 Apr 2022

%\cite{}
\bibitem{Ukwatta:2009xk}
T.~N.~Ukwatta, \textit{et al.},
%J.~H.~MacGibbon, W.~C.~Parke, K.~S.~Dhuga, A.~Eskandarian, N.~Gehrels, L.~Maximon and D.~C.~Morris,
%``Spectral Lags of Gamma-Ray Bursts from Primordial Black Hole (PBH) Evaporations,''
AIP Conf. Proc. \textbf{1133}, no.1, 440-442 (2009)
%%doi:10.1063/1.3155947
[arXiv:0901.0542 [astro-ph.HE]].
%5 citations counted in INSPIRE as of 28 Apr 2022

\bibitem{MacGibbon:2015mya}
J.~H.~MacGibbon, T.~N.~Ukwatta, J.~T.~Linnemann, S.~S.~Marinelli, D.~Stump and K.~Tollefson,
%``Primordial Black Holes,''
[arXiv:1503.01166 [astro-ph.HE]].
%8 citations counted in INSPIRE as of 07 Jul 2022

   %\cite{}
\bibitem{Page:1976df}
D.~N.~Page,
%``Particle Emission Rates from a Black Hole: Massless Particles from an Uncharged, Nonrotating Hole,''
Phys. Rev. D\textbf{13}, 198-206 (1976).
%%doi:10.1103/PhysRevD.13.198
%884 citations counted in INSPIRE as of 23 Aug 2021
  
  
\bibitem{Page:1976ki}
D.~N.~Page,
%``Particle Emission Rates from a Black Hole. 2. Massless Particles from a Rotating Hole,''
Phys. Rev. D\textbf{14}, 3260-3273 (1976).
%%doi:10.1103/PhysRevD.14.3260
%366 citations counted in INSPIRE as of 23 Aug 2021

%\cite{Chambers:1997ai, Taylor:1998dk}
\bibitem{Chambers:1997ai}
C.~M.~Chambers, W.~A.~Hiscock and B.~Taylor,
%``Spinning down a black hole with scalar fields,''
Phys. Rev. Lett. \textbf{78}, 3249-3251 (1997).
%%doi:10.1103/PhysRevLett.78.3249
%[arXiv:gr-qc/9703018 [gr-qc]].
%21 citations counted in INSPIRE as of 23 Aug 2021

%\cite{Taylor:1998dk}
\bibitem{Taylor:1998dk}
B.~E.~Taylor, C.~M.~Chambers and W.~A.~Hiscock,
%``Evaporation of a Kerr black hole by emission of scalar and higher spin particles,''
Phys. Rev. D\textbf{58}, 044012 (1998).
%%doi:10.1103/PhysRevD.58.044012
%[arXiv:gr-qc/9801044 [gr-qc]].
%34 citations counted in INSPIRE as of 23 Aug 2021

\bibitem{MacGibbon:1990zk}
J.~H.~MacGibbon and B.~R.~Webber,
%``Quark and gluon jet emission from primordial black holes: The instantaneous spectra,''
Phys. Rev. D\textbf{41}, 3052-3079 (1990).
%%doi:10.1103/PhysRevD.41.3052
%223 citations counted in INSPIRE as of 23 Aug 2021

%\cite{}
\bibitem{MacGibbon:1991tj}
J.~H.~MacGibbon,
%``Quark and gluon jet emission from primordial black holes. 2. The Lifetime emission,''
Phys. Rev. D \textbf{44}, 376-392 (1991)
%doi:10.1103/PhysRevD.44.376
%169 citations counted in INSPIRE as of 28 Apr 2022


%\cite{}
\bibitem{MacGibbon:1991vc}
J.~H.~MacGibbon and B.~J.~Carr,
%``Cosmic rays from primordial black holes,''
Astrophys. J. \textbf{371}, 447-469 (1991).
%%doi:10.1086/169909
%216 citations counted in INSPIRE as of 28 Apr 2022

%\cite{}
\bibitem{MacGibbon:2007yq}
J.~H.~MacGibbon, B.~J.~Carr and D.~N.~Page,
%``Do Evaporating Black Holes Form Photospheres?,''
Phys. Rev. D \textbf{78}, 064043 (2008)
%doi:10.1103/PhysRevD.78.064043
[arXiv:0709.2380 [astro-ph]].
%64 citations counted in INSPIRE as of 07 Jul 2022


%\cite{MacGibbon:1991vc, MacGibbon:2007yq, MacGibbon:2010nt}
\bibitem{MacGibbon:2010nt}
J.~H.~MacGibbon, B.~J.~Carr and D.~N.~Page,
%``Do Evaporating 4D Black Holes Form Photospheres and/or Chromospheres?,''
%doi:10.1142/9789814374552\_0157
[arXiv:1003.3901 [astro-ph.HE]].
%3 citations counted in INSPIRE as of 07 Jul 2022

%\cite{}
\bibitem{Ukwatta:2015iba}
T.~N.~Ukwatta, D.~R.~Stump, J.~T.~Linnemann, J.~H.~MacGibbon, S.~S.~Marinelli, T.~Yapici and K.~Tollefson,
%``Primordial Black Holes: Observational Characteristics of The Final Evaporation,''
Astropart. Phys. \textbf{80}, 90-114 (2016)
%%doi:10.1016/j.astropartphys.2016.03.007
%[arXiv:1510.04372 [astro-ph.HE]].
%32 citations counted in INSPIRE as of 28 Apr 2022


\bibitem{Calza:2021czr}
M.~Calz\`a, J.~March-Russell and J.~G.~Rosa,
%``Evaporating primordial black holes, the string axiverse, and hot dark radiation,''
[arXiv:2110.13602 [astro-ph.CO]].

\bibitem{Arvanitaki:2009fg}
A.~Arvanitaki, S.~Dimopoulos, S.~Dubovsky, N.~Kaloper and J.~March-Russell,
%``String Axiverse,''
Phys. Rev. D \textbf{81}, 123530 (2010)
%doi:10.1103/PhysRevD.81.123530
[arXiv:0905.4720 [hep-th]].
%1285 citations counted in INSPIRE as of 07 Jul 2022

%\cite{Peccei:1977hh, Wilczek:1977pj}
\bibitem{Peccei:1977hh}
R.~D.~Peccei and H.~R.~Quinn,
%``CP Conservation in the Presence of Instantons,''
Phys. Rev. Lett. \textbf{38}, 1440-1443 (1977).
%doi:10.1103/PhysRevLett.38.1440
%6714 citations counted in INSPIRE as of 12 Oct 2022

%\cite{Wilczek:1977pj}
\bibitem{Wilczek:1977pj}
F.~Wilczek,
%``Problem of Strong  $P$  and  $T$  Invariance in the Presence of Instantons,''
Phys. Rev. Lett. \textbf{40}, 279-282 (1978).
%doi:10.1103/PhysRevLett.40.279
%4549 citations counted in INSPIRE as of 12 Oct 2022

%\cite{Baker:2021btk, Baker:2022rkn}
\bibitem{Baker:2021btk}
M.~J.~Baker and A.~Thamm,
%``Probing the particle spectrum of nature with evaporating black holes,''
SciPost Phys. \textbf{12}, no.5, 150 (2022)
%doi:10.21468/SciPostPhys.12.5.150
[arXiv:2105.10506 [hep-ph]].
%15 citations counted in INSPIRE as of 07 Oct 2022

%\cite{Baker:2022rkn}
\bibitem{Baker:2022rkn}
M.~J.~Baker and A.~Thamm,
%``Black Hole Evaporation Beyond the Standard Model of Particle Physics,''
[arXiv:2210.02805 [hep-ph]].
%0 citations counted in INSPIRE as of 07 Oct 2022
  

%
\bibitem{Arbey:2019mbc}
A.~Arbey and J.~Auffinger,
%``BlackHawk: A public code for calculating the Hawking evaporation spectra of any black hole distribution,''
Eur. Phys. J. C \textbf{79}, no.8, 693 (2019)
%doi:10.1140/epjc/s10052-019-7161-1
[arXiv:1905.04268 [gr-qc]].
%69 citations counted in INSPIRE as of 07 Jul 2022

\bibitem{Arbey:2020yzj}
A.~Arbey, J.~Auffinger and J.~Silk,
%``Primordial Kerr Black Holes,''
PoS \textbf{ICHEP2020}, 585 (2021)
%doi:10.22323/1.390.0585
[arXiv:2012.14767 [astro-ph.CO]].
%2 citations counted in INSPIRE as of 12 Jul 2022

%\cite{}
\bibitem{Arbey:2021yke}
A.~Arbey, J.~Auffinger, M.~Geiller, E.~R.~Livine and F.~Sartini,
%``Hawking radiation by spherically-symmetric static black holes for all spins. II. Numerical emission rates, analytical limits, and new constraints,''
Phys. Rev. D \textbf{104}, no.8, 084016 (2021)
%doi:10.1103/PhysRevD.104.084016
[arXiv:2107.03293 [gr-qc]].
%8 citations counted in INSPIRE as of 07 Jul 2022

%\cite{}
\bibitem{Arbey:2021mbl}
A.~Arbey and J.~Auffinger,
%``Physics Beyond the Standard Model with BlackHawk v2.0,''
Eur. Phys. J. C \textbf{81}, 10 (2021)
%doi:10.1140/epjc/s10052-021-09702-8
[arXiv:2108.02737 [gr-qc]].
%21 citations counted in INSPIRE as of 07 Jul 2022


%
\bibitem{Sjostrand:2007gs}
T.~Sjostrand, S.~Mrenna and P.~Z.~Skands,
%``A Brief Introduction to PYTHIA 8.1,''
Comput. Phys. Commun. \textbf{178}, 852-867 (2008)
%doi:10.1016/j.cpc.2008.01.036
[arXiv:0710.3820 [hep-ph]].
%6521 citations counted in INSPIRE as of 07 Jul 2022

%\cite{}
\bibitem{Bierlich:2022pfr}
C.~Bierlich, %S.~Chakraborty, N.~Desai, L.~Gellersen, I.~Helenius, P.~Ilten, L.~L\"onnblad, S.~Mrenna, S.~Prestel and C.~T.~Preuss,
 \textit{et al.}
%``A comprehensive guide to the physics and usage of PYTHIA 8.3,''
[arXiv:2203.11601 [hep-ph]].
%12 citations counted in INSPIRE as of 07 Jul 2022


\bibitem{Coogan:2019qpu}
A.~Coogan, L.~Morrison and S.~Profumo,
%``Hazma: A Python Toolkit for Studying Indirect Detection of Sub-GeV Dark Matter,''
JCAP \textbf{01}, 056 (2020)
%doi:10.1088/1475-7516/2020/01/056
[arXiv:1907.11846 [hep-ph]].
%30 citations counted in INSPIRE as of 08 Jul 2022

\bibitem{Coogan:2020tuf}
A.~Coogan, L.~Morrison and S.~Profumo,
%``Direct Detection of Hawking Radiation from Asteroid-Mass Primordial Black Holes,''
Phys. Rev. Lett. \textbf{126}, no.17, 171101 (2021)
%doi:10.1103/PhysRevLett.126.171101
[arXiv:2010.04797 [astro-ph.CO]].
%45 citations counted in INSPIRE as of 28 Apr 2022
  
%\cite{Teukolsky:1972my, Teukolsky:1973, Press:1973zz, Teukolsky:1974yv}
\bibitem{Teukolsky:1972my}
S.~A.~Teukolsky,
%``Rotating black holes - separable wave equations for gravitational and electromagnetic perturbations,''
Phys. Rev. Lett. \textbf{29}, 1114-1118 (1972).
%doi:10.1103/PhysRevLett.29.1114
%717 citations counted in INSPIRE as of 07 Jul 2022

  %\cite{}
\bibitem{Teukolsky:1973} 
  S.~A.~Teukolsky,
  %``Perturbations of a rotating black hole. I - Fundamental equations for gravitational, electromagnetic, and neutrino-field perturbations,''
  Astrophys.\ J.\  {\bf 185}, 635 (1973).
   
  %\cite{}
\bibitem{Press:1973zz} 
  W.~H.~Press and S.~A.~Teukolsky,
  %``Perturbations of a Rotating Black Hole. II. Dynamical Stability of the Kerr Metric,''
  Astrophys.\ J.\  {\bf 185}, 649 (1973).
  %%CITATION = ASJOA,185,649;%%
  %245 citations counted in INSPIRE as of 24 Oct 2014

\bibitem{Teukolsky:1974yv} 
  S.~A.~Teukolsky and W.~H.~Press,
  %``Perturbations of a rotating black hole. III - Interaction of the hole with gravitational and electromagnetic radiation,''
  Astrophys.\ J.\  {\bf 193}, 443 (1974).
  %%CITATION = ASJOA,193,443;%%
  %265 citations counted in INSPIRE as of 22 Mar 2014
  
 
\bibitem{Rosa:2016bli}
J.~G.~Rosa,
%``Superradiance in the sky,''
Phys. Rev. D \textbf{95}, no.6, 064017 (2017)
%doi:10.1103/PhysRevD.95.064017
[arXiv:1612.01826 [gr-qc]].
%13 citations counted in INSPIRE as of 08 Jul 2022. 
  
 %\cite{}
\bibitem{Altarelli:1977zs}
G.~Altarelli and G.~Parisi,
%``Asymptotic Freedom in Parton Language,''
Nucl. Phys. B \textbf{126}, 298-318 (1977).
%doi:10.1016/0550-3213(77)90384-4
%7670 citations counted in INSPIRE as of 28 Apr 2022


\bibitem{Chen:2016wkt}
J.~Chen, T.~Han and B.~Tweedie,
%``Electroweak Splitting Functions and High Energy Showering,''
JHEP \textbf{11}, 093 (2017)
%doi:10.1007/JHEP11(2017)093
[arXiv:1611.00788 [hep-ph]].
%67 citations counted in INSPIRE as of 28 Apr 2022 
  


%%%%%%%%%%%%%%%%%%%%%%%%%%%%%%%%%%%%%%%%%%%%%%%%%%%%%%%%%%%



%\cite{AMEGO:2019gny}
\bibitem{AMEGO:2019gny}
R.~Caputo \textit{et al.} [AMEGO],
%``All-sky Medium Energy Gamma-ray Observatory: Exploring the Extreme Multimessenger Universe,''
[arXiv:1907.07558 [astro-ph.IM]].
%80 citations counted in INSPIRE as of 28 Sep 2022

%\cite{Fleischhack:2021mhc}
\bibitem{Fleischhack:2021mhc}
H.~Fleischhack,
%``AMEGO-X: MeV gamma-ray Astronomy in the Multi-messenger Era,''
PoS \textbf{ICRC2021}, 649 (2021)
%doi:10.22323/1.395.0649
[arXiv:2108.02860 [astro-ph.IM]].
%19 citations counted in INSPIRE as of 28 Sep 2022

%\cite{e-ASTROGAM:2016bph}
\bibitem{e-ASTROGAM:2016bph}
A.~De Angelis \textit{et al.} [e-ASTROGAM],
%``The e-ASTROGAM mission,''
Exper. Astron. \textbf{44}, no.1, 25-82 (2017)
%doi:10.1007/s10686-017-9533-6
[arXiv:1611.02232 [astro-ph.HE]].
%150 citations counted in INSPIRE as of 28 Sep 2022

%\cite{Tatischeff:2019mun}
\bibitem{Tatischeff:2019mun}
V.~Tatischeff, %A.~De Angelis, M.~Tavani, U.~Oberlack, R.~Walter, G.~Ambrosi, A.~Argan, P.~von Ballmoos, S.~Brandt and A.~Bulgarelli, 
\textit{et al.}
%``All-Sky-ASTROGAM: The MeV Gamma-Ray Companion to Multimessenger Astronomy,''
Mem. Soc. Ast. It. \textbf{90}, no.1-2, 137-143 (2019)
[arXiv:1905.07806 [astro-ph.HE]].
%3 citations counted in INSPIRE as of 28 Sep 2022


%\cite{Dzhatdoev:2019kay}
\bibitem{Dzhatdoev:2019kay}
T.~Dzhatdoev and E.~Podlesnyi,
%``Massive Argon Space Telescope (MAST): A concept of heavy time projection chamber for $\gamma$-ray astronomy in the 100 MeV\textendash{}1 TeV energy range,''
Astropart. Phys. \textbf{112}, 1-7 (2019)
%doi:10.1016/j.astropartphys.2019.04.004
[arXiv:1902.01491 [astro-ph.HE]].
%15 citations counted in INSPIRE as of 28 Sep 2022

%\cite{Halzen:1995hu, Dave:2019epr, Capanema:2021hnm, Bernal:2022swt}
\bibitem{Halzen:1995hu}
F.~Halzen, B.~Keszthelyi and E.~Zas,
%``Neutrinos from primordial black holes,''
Phys. Rev. D \textbf{52}, 3239-3247 (1995)
%doi:10.1103/PhysRevD.52.3239
[arXiv:hep-ph/9502268 [hep-ph]].
%39 citations counted in INSPIRE as of 07 Oct 2022

%\cite{Dave:2019epr}
\bibitem{Dave:2019epr}
P.~Dave \textit{et al.} [IceCube],
%``Neutrinos from Primordial Black Hole Evaporation,''
PoS \textbf{ICRC2019}, 863 (2021)
%doi:10.22323/1.358.0863
[arXiv:1908.05403 [astro-ph.HE]].
%9 citations counted in INSPIRE as of 07 Oct 2022


%\cite{Capanema:2021hnm}
\bibitem{Capanema:2021hnm}
A.~Capanema, A.~Esmaeili and A.~Esmaili,
%``Evaporating primordial black holes in gamma ray and neutrino telescopes,''
JCAP \textbf{12}, no.12, 051 (2021)
%doi:10.1088/1475-7516/2021/12/051
[arXiv:2110.05637 [hep-ph]].
%9 citations counted in INSPIRE as of 07 Oct 2022

%\cite{Bernal:2022swt}
\bibitem{Bernal:2022swt}
N.~Bernal, V.~Mu\~noz-Albornoz, S.~Palomares-Ruiz and P.~Villanueva-Domingo,
%``Current and future neutrino limits on the abundance of primordial black holes,''
[arXiv:2203.14979 [hep-ph]].
%5 citations counted in INSPIRE as of 07 Oct 2022

\end{thebibliography}
\end{document}